\documentclass[showpacs,aps,prd,nofootinbib,floatfix,amsmath,amssymb]{revtex4}
\usepackage{graphicx}
\begin{document}

\makeatletter
\newbox\slashbox \setbox\slashbox=\hbox{$/$}
\newbox\Slashbox \setbox\Slashbox=\hbox{\large$/$}
\def\pFMslash#1{\setbox\@tempboxa=\hbox{$#1$}
  \@tempdima=0.5\wd\slashbox \advance\@tempdima 0.5\wd\@tempboxa
  \copy\slashbox \kern-\@tempdima \box\@tempboxa}
\def\pFMSlash#1{\setbox\@tempboxa=\hbox{$#1$}
  \@tempdima=0.5\wd\Slashbox \advance\@tempdima 0.5\wd\@tempboxa
  \copy\Slashbox \kern-\@tempdima \box\@tempboxa}
\def\FMslash{\protect\pFMslash}
\def\FMSlash{\protect\pFMSlash}
\def\miss#1{\ifmmode{/\mkern-11mu #1}\else{${/\mkern-11mu #1}$}\fi}
\makeatother

\title{Decays $Z'\to \gamma \gamma \gamma$ and $Z\to \gamma \gamma \gamma$ in the minimal $331$ model}
\author{J. Monta\~no$^{(a)}$, M. A. P\' erez$^{(a)}$, F. Ram\'\i rez-Zavaleta$^{(b)}$, and J. J. Toscano$^{(c)}$}
\address{$^{(a)}$ Departamento de F\'\i sica, CINVESTAV.
Apartado Postal 14-740, 07000, M\'exico D.F., M\'exico.\\
$^{(b)}$Facultad de Ciencias F\'\i sico Matem\'aticas,
Universidad Michoacana de San Nicol\'as de
Hidalgo, Avenida Francisco J. M\'ujica S/N, 58060, Morelia, Michoac\'an, M\'exico. \\
$^{(c)}$Facultad de Ciencias F\'isico Matem\'aticas,
Benem\'erita Universidad Aut\'onoma de Puebla, Apartado Postal
1152, Puebla, Puebla, M\'exico.}
\begin{abstract}
 The possibility of a significant effect of exotic particles on the $Z'\to \gamma \gamma \gamma$ and $Z\to \gamma \gamma \gamma$ decays is investigated in the context of the minimal 331 model. This model, which is based in the $SU_C(3)\times SU_L(3)\times U_X(1)$ gauge group, predicts the existence of many exotic charged particles that can significantly enhance the decay widths. It is found that the standard model prediction for the $Z\to \gamma \gamma \gamma$ decay remains essentially unchanged, as the new physics effects quickly decouples. On the other hand, it is found that the contributions of the new exotic quarks and gauge bosons predicted by this model lead to a branching fraction for the $Z'\to \gamma \gamma \gamma$ decay of about $10^{-6}$, which is about 3 orders of magnitude larger than that of the $Z\to \gamma \gamma \gamma$ decay.
\end{abstract}


\maketitle

\section{Introduction}
\label{in}
Numerous extensions of the Standard Model (SM) predict the existence of new neutral gauge bosons $Z'$~\cite{Langacker1}. This class of gauge bosons can be associated with spontaneous symmetry breaking (SSB) of additional $U(1)s$ gauge groups or with Kaluza-Klein excitations of theories with extra compact  dimensions~\cite{ED}. Phenomenologically, the most interesting option is the breaking of these $U(1)s$ at around TeV scales, giving rise to extra neutral gauge bosons observable at the Large Hadron Collider (LHC). Although the LHC has been designed to solve the Higgs puzzle, it is possible that in the first stage of its running signals of new physics show up, such as the resonance produced by a new $Z'$ gauge boson decaying into charged leptons~\cite{Rizzo1}. Therefore, it is worth investigating the phenomenology of this particle. In this paper, we will focus on the $Z'$ gauge boson predicted by the minimal $331$ model~\cite{P331,F331}, which predicts new physics at the TeV scale. In particular, we are interested in studying the rare decay of this $Z'$ gauge boson into three photons. These types of decays are naturally suppressed in renormalizable theories, as they first arise at the one-loop level. The analogous decay in the SM $Z\to \gamma \gamma \gamma$ is quite suppressed, with a decay width of the order of $10^{-10}$ GeV~\cite{Z3gFermionA,Z3gFermionWE,Z3gW}. In the SM, the fermionic contribution was calculated in an approximate way more than two decades ago~\cite{Z3gFermionA} and some years after the complete analytical results were presented~\cite{Z3gFermionWE}. As the $W$ gauge boson contribution is concerned, it was presented in~\cite{Z3gFermionWE,Z3gW}. It was shown in~\cite{Z3gFermionWE} that the fermionic-$W$ interference is significant. Since there are no charged scalars particles in the SM this kind of contribution only occurs in models with extended Higgs sectors. In~\cite{Z3gH} such a contribution is analyzed in the context of the popular two Higgs doublet model and the minimal supersymmetric standard model. It was found that this contribution is several orders of magnitude lower than the fermionic and $W$ contributions. A more general analysis of the contribution of all three types of particles to this decay was presented in~\cite{SuperGraph}.

As already commented, we are interested in studying the $Z'\to \gamma \gamma \gamma$ decay in the context of the minimal $331$ model~\cite{P331,F331}. This model, which is based in the $SU_C(3)\times SU_L(3)\times U_X(1)$ gauge group, has attracted the attention of numerous authors in the past decade, mainly because it possesses some peculiar features that are not  present in other SM extensions. Its more interesting property is that anomalies do not cancel in each generation independently as in the SM, but such a cancelation occurs only when the three generations are taken into account together. The cancellation mechanism requires that the number of families must be an integer multiple of the color number. This fact, together with the property of asymptotic freedom of QCD, which establishes that the color number is less than five, implies that the 331 model predicts the existence of only three fermionic families. Another interesting property of the minimal version of this model is that the weak angle is subject to the constraint $s^2_W<\frac{1}{4}$~\cite{P331}\footnote{From now on, $s_W$ and $c_W$ stand for sine and cosine of the weak angle $\theta_W$.}. It results that, when it is evolved to high values, the model loses its perturbative character at a scale about $8$ TeV~\cite{LP}. The fact that the value of $s^2_W$ is very close to $1/4$ leads to an upper bound on the scale associated with the first stage of SSB, when the  $SU_C(3)\times SU_L(3)\times U_X(1)$ group is broken down into the SM group  $SU_C(3)\times SU_L(2)\times U_Y(1)$, which translates directly into the gauge bosons that acquire masses at this scale, among them the $Z'$ gauge boson~\cite{F331,Ng1}. Then, the 331 model is phenomenologically well motivated to be probed in the LHC.

In the minimal 331 model the lepton spectrum is the same as in the SM, but it is arranged in antitriplets of the gauge group $SU_L(3)$. The quark sector is also arranged in the fundamental representation of this group, which requires the introduction of three new quarks. Along with the $Z'$ boson, four additional charged gauge bosons are predicted by the minimal 331 model: two singly charged bosons $Y^\pm$ and two doubly charged ones $Y^{\pm \pm}$. These gauge bosons carry two units of lepton number and so have been classified as bileptons~\cite{Bileptons}. The new gauge bosons together with the exotic quarks are endowed with mass at the first stage of SSB, when $SU_L(3)\times U_X(1)$ is broken into $SU_L(2)\times U_Y(1)$~\cite{MassesYs}. Since $SU_L(2)$ is completely embedded into $SU_L(3)$, the couplings between the SM and the extra gauge bosons are determined by the coupling constant $g$ associated with the $SU_L(2)$ group and the weak angle $\theta_W$~\cite{MassesYs,T1}. On the other hand, the Higgs sector of the minimal 331 model consists of three triplet and one sextet, but only one triplet is needed to break down $SU_L(3)\times U_X(1)$ into $SU_L(2)\times U_Y(1)$. At this stage, the Higgs sector consists of three doublets and one triplet of $SU_L(2)$, as well as of four complex singlets~\cite{Ng1,T1}. After breaking the usual electroweak group into the electromagnetic one, the physical scalar sector is composed by: five neutral CP-even $H_i\, (i=1,\cdots , 5)$, three neutral CP-odd $A_i\, (i=1,2,3)$, four charged $h^\pm_i\, (i=1,\cdots, 4)$, and three doubly charged $d^{\pm \pm}_i\, (i=1,2 ,3)$, from which only $H_1$ is light, with mass of the order of the Fermi scale $v$~\cite{ScalarS}. In a previous communication by some of us~\cite{T2}, a comprehensive analysis of tree-level two-body decays of $Z'$, including the one-loop induced ones $Z'\to Z\gamma$ and $Z'\to ZZ$, was performed in the context of the minimal 331 model. Our main goal in this work is to investigate the sensitivity of the $V^0\to \gamma \gamma \gamma$ $(V^0=Z,Z')$ decays to virtual effects of new particles of spin $0$, $1/2$, and $1$ living at the TeV scale. As already commented, one interesting peculiarity of the $331$ model is the presence of new particles with a charge content that differs from that carried by the known particles. Exotic particles with a charge content $Q>1$ in units of the positron charge may enhance substantially the amplitude of the $V^0\gamma \gamma \gamma$ vertex, as it is proportional to $Q^3$, which in turns leads to a probability proportional to $Q^6$. This class of effect was studied by some of us some years ago in light by light scattering, showing that this process is quite sensitive to the contribution of the doubly charged bileptons, as they modify the cross section by an additional factor of $Q^8=2^8$~\cite{T3}. In some more recent communications by some of us, the impact of the exotic quarks on the rare $V^0\to ggg$~\cite{T4} and $V^0\to gg\gamma$~\cite{T5} decays was studied. All the new charged particles that predict the model receive their mass at the first stage of SSB, so that a good strategy to estimate the impact of scalars, fermions, and vectors to the $V^0\to \gamma \gamma \gamma$ decay is to assume that each set of these three types of particles are mass degenerate. Accordingly, the impact of the three exotic quarks, two of them ($D$ and $S$) with charge content of $Q_D=Q_S=-4/3$, and the third one ($T$) with charge $Q_T=5/3$, is $Q^6_D+Q^6_S+Q^6_T\approx 32.7$. On the other hand, the impact of the simple and doubly charged bileptons is $Q^6_{Y^{\pm \pm}}+Q^6_{Y^\pm}=65$. As far as the four simply charged and three doubly charged Higgs bosons is concerned, their impact would be proportional to $3Q^6_{d^{\pm \pm}}+4 Q^6_{h^\pm}=196$.

The rest of the paper has been organized as follows. In Sec. \ref{m} a brief discussion of the $331$ model is presented. Section \ref{a} is devoted to present compact analytical expressions for the amplitudes associated with the $V^0\to \gamma \gamma \gamma$ decay, with $V^0=Z, Z'$. In Sec. \ref{rd}, we discuss our results. Finally, in Sec. \ref{c} the conclusions are presented.

\section{The minimal $331$ model}
\label{m} As already mentioned in the introduction, the lepton spectrum of the model is the same as in the SM and are accommodated as antitriplets of $SU_L(3)$:
\begin{equation}
L_{1,2,3}=\left(\begin{array}{ccc}
e \\
\, \, \\
\nu_e \\
\, \, \\
e^c
\end{array}\right)_L, \, \,
\left(\begin{array}{ccc}
\mu \\
\, \, \\
\nu_\mu \\
\, \, \\
\mu^c
\end{array}\right)_L, \, \,
\left(\begin{array}{ccc}
\tau \\
\, \, \\
\nu_\tau \\
\, \, \\
\tau^c
\end{array}\right)_L, \, \,
(1,3^*,0)\, .
\end{equation}
Notice that $(\ell^c_a)_L=(\ell_{aR})^c$ ($\ell_a=e,\mu,\tau$). In order to cancel the $SU_L(3)$ anomaly, the same number of fermion triplets and antitriplets is necessary. This requires one to arrange two quark generations as triplets and the other one as an antitriplet. It is customary to choose the third generation as the one transforming as antitriplet in order to distinguish the new dynamic effects in the physics of the quark top from that of the lighter generations. Accordingly, the three generations are specified as follows:
\begin{equation}
Q_{1,2}=\left(\begin{array}{ccc}
u \\
\, \, \\
d \\
\, \, \\
D
\end{array}\right)_L, \, \,
\left(\begin{array}{ccc}
c \\
\, \, \\
s \\
\, \, \\
S
\end{array}\right)_L, \, \, (3,3,-1/3)\, ,
\end{equation}
\begin{equation}
Q_{3}=\left(\begin{array}{ccc}
b \\
\, \, \\
t \\
\, \, \\
T
\end{array}\right)_L, \, \,
 (3,3^*,2/3)\, .
\end{equation}
On the other hand, the right-handed quarks are specified as follows:
\begin{equation}
d^c, \, \, s^c, \, \, b^c\, : \, (3^*,1,1/3) , \, \, \, D^c, \, \, S^c\, : \, (3^*,1,4/3)\, ,
\end{equation}
\begin{equation}
u^c, \, \, c^c, \, \, t^c\, : \, (3^*,1,-2/3) , \, \, \, T^c\, : \, (3^*,1,-5/3)\, .
\end{equation}
In the first stage of SSB, when the $SU_L(3)\times U_X(1)$ group is broken into the usual electroweak group $SU_L(2)\times U_Y(1)$, only the three new quarks $D,S$, and $T$ acquire masses. These exotic quarks arise as singlets of the $SU_L(2)$ group, so they do not couple to the $W$ gauge boson. However, they do couple to both the $Z$ and $Z'$ gauge bosons~\cite{T2}.

On the other hand, as already commented in the introduction, the Higgs sector of the minimal 331 model is comprised of three triplets and one sextet of $SU_L(3)$: \begin{equation}
\phi_Y=\left(\begin{array}{ccc}
\Phi_Y \\
\, \, \\
\phi^0 \\
\end{array}\right): \, \,
 (1,3,1)\, , \, \,
 \phi_1=\left(\begin{array}{ccc}
\Phi_1 \\
\, \, \\
\delta^- \\
\end{array}\right): \, \,
 (1,3,0)\, , \, \,
 \phi_2=\left(\begin{array}{ccc}
\tilde{\Phi}_2 \\
\, \, \\
\rho^{--} \\
\end{array}\right): \, \,
 (1,3,-1)\, ,
\end{equation}

\begin{equation}
H=\left(\begin{array}{ccc}
T & \tilde{\Phi}_3/\sqrt{2} \\
\, \, \\
\tilde{\Phi}^T_3/\sqrt{2} & \eta^{--}
\end{array}\right): \, \,
 (1,6,0)\, .
\end{equation}
To break $SU_L(3)\times U_X(1)$ into $SU_L(2)\times U_Y(1)$, only the $\phi_Y$ scalar triplet of $SU_L(3)$ is required. The hypercharge is identified as a linear combination of the broken generators $T^8$ and $X$: $Y=\sqrt{3}(\lambda^8+\sqrt{2}X\lambda^9)$, with $\lambda^8$ a Gell-Mann matrix and $\lambda^9=\sqrt{2/3} \,\mathrm{diag}(1,1,1)$. The next stage of SSB occurs at the Fermi scale and is achieved by the two triplets $\phi_1$ and
$\phi_2$. The sextet $H$ is necessary to provide realistic masses for the leptons~\cite{Trip}. In these expressions $\Phi_Y$ , $\Phi_1$ , $\tilde{\Phi}_2=i\sigma^2\Phi^*_2$ and $\Phi_3$ are all doublets of $SU_L(2)$ with hypercharge 3, 1, 1, and 1, respectively. On the other hand, $T$ is an
$SU_L(2)$ triplet with $Y=+2$, whereas $\delta^-$, $\rho^{--}$, and $\eta^{--}$
are all singlets of $SU_L(2)$ with hypercharge $-2$ $-4$, and $+4$, respectively~\cite{MassesYs}. The extra $Z'$ boson, the bileptons
and the exotic quarks get masses at the first stage of SSB through the vacuum expectation value $<\phi_Y>_0=(0,0,u/\sqrt{2})$. The bileptons form a $SU_L(2)$ doublet with hypercharge $+3$. The spectrum of physical gauge particles is the following. The charged gauge bosons are given by
\begin{eqnarray}
Y^{++}_\mu&=&\frac{1}{\sqrt{2}}(A^4_\mu -iA^5_\mu) \, ,\\
Y^{+}_\mu&=&\frac{1}{\sqrt{2}}(A^6_\mu -iA^7_\mu) \, ,
\end{eqnarray}

\begin{equation}
W^{+}_\mu=\frac{1}{\sqrt{2}}(A^1_\mu -iA^2_\mu) \, ,
\end{equation}
with $m^2_{Y^{++}}=g^2(u^2+v^2_2+3v^2_3)/4$, $m^2_{Y^{+}}=g^2(u^2+v^2_1+v^2_3)/4$, and $m^2_{W}=g^2(v^2_1+v^2_2+v^2_3)/4$. The hierarchy of the SSB yields a splitting between the bilepton masses given by $|m^2_{Y^{+}}-m^2_{Y^{++}}|\leq 3m^2_{W}$. However, to simplify the discussion we will consider only the degenerate case.

In the neutral sector, the gauge fields $(A^3,A^8,X)$ define three mass eigenstates $(A,Z_1 ,Z_2)$ via the following rotation:
\begin{equation}
\left(\begin{array}{ccc}
A_\mu \\
\, \\
Z_\mu \\
\, \\
Z'_\mu
\end{array}\right)
=\left(\begin{array}{ccc}
s_W & \sqrt{3}s_W &\sqrt{1-4s^2_W} \\
\, \, \\
c_W & -\sqrt{3}s_Wt_W & -t_W\sqrt{1-4s^2_W} \\
\, \\
0 &-\frac{\sqrt{1-4s^2_W}}{c_W} & \sqrt{3}t_W
\end{array}\right)\left(\begin{array}{ccc}
A^3_\mu \\
\, \\
A^8_\mu \\
\, \\
X_\mu
\end{array}\right) \, ,
\end{equation}

\begin{equation}
\left(\begin{array}{ccc}
Z_{1\mu} \\
\, \\
Z_{2\mu}
\end{array}\right)
=\left(\begin{array}{ccc}
\cos\theta & -\sin\theta\\
\, \, \\
\sin\theta & \cos\theta
\end{array}\right)\left(\begin{array}{ccc}
Z_\mu \\
\, \\
Z'_\mu
\end{array}\right) \, ,
\end{equation}
where the mixing angle is
\begin{equation}
\sin^2\theta=\frac{m^2_Z-m^2_{Z_1}}{m^2_{Z_2}-m^2_{Z_1}}\, ,
\end{equation}
with $m^2_Z=m^2_W/c^2_W$ and $Z_1$ standing for the SM $Z$ boson. In this paper, we will work in the approximation $\theta=0$, so $Z_1$ and $Z_2$ coincide with $Z$ and $Z'$, respectively.

To calculate the amplitudes for the $V^0\to \gamma \gamma \gamma$  decay, we need the Feynman rules for the couplings of $Z$ and $Z'$ to all the charged particles of the model. The couplings of  $Z$ and $Z'$ to leptons and quarks, including the exotic ones, are all given in~\cite{T2} and we refrain from including them here. The couplings of these particles with photons are dictated by spinorial electrodynamics. As far as the couplings of the neutral gauge bosons $A$, $Z$, and $Z'$ with the charged ones $W^+$, $Y^+$, and $Y^{++}$, they depend on the gauge-fixing procedure used to quantize the theory. The calculation of these contributions are  greatly simplified if one uses a nonlinear gauge-fixing procedure. To carry out the $W^+$ contribution we used a covariant gauge-fixing procedure as the one presented in~\cite{HT,T1}. As far as the bilepton contribution is concerned, we used the Feynman rules that arise from the gauge-fixing procedure used in~\cite{T6,T7}. The Feynman rules for all possible couplings among the neutral and charged gauge bosons of the minimal 331 model can be found at the Appendix~\ref{FR}. Some of these gauge bosons rules have been worked in \cite{T1,T2,T6}. On the other hand, the coupling of charged scalars to photons are model independent, as they are dictated by scalar electrodynamics. Consequently, we only need the couplings of all the charged scalars with the $Z$ and $Z'$ gauge bosons. These couplings arise from the Higgs kinetic sectors of the 331 model, but to determine them one needs to diagonalize the Higgs potential of the model. We have used the diagonalization given in~\cite{ScalarS} to determine all couplings of the $Z$ and $Z'$ gauge bosons with the charged scalars of the model. The vertex functions associated with the $V^{0\alpha} S^\dag (k_1)S(k_2)$, $V^{0\alpha} A^\beta S^\dag S$, and $A^\alpha A^\beta S^\dag S$ couplings, with $S$ stands for a charged scalar, are given, respectively, by
\begin{equation}
\frac{igg^S_{V^0}}{2c_W}(k_1-k_2)_\alpha \quad , \quad
\frac{ieQ_Sgg^S_{V^0}}{c_W}g_{\alpha \beta} \quad , \quad
i2e^2Q_S^2g_{\alpha \beta} \quad ,
\end{equation}
where all momentum are taken incoming to the vertex. The values of $Q_S$ and $g^S_V$ are presented in Table \ref{Table1}. In this table, $\theta$ is the $Z-Z'$ mixing angle, which we have taken equal to zero. On the other hand, the $H_{22}$, $H_{32}$, $\tilde{a}$, $\tilde{N}_4$, $\tilde{N}_5$, $\tilde{X}_4$, and $\tilde{X}_5$ are given by
\begin{equation}\label{}
    H_{22}=-\frac{1}{\sqrt{2}} \quad , \quad
    H_{32}=-\frac{\lambda}{\tilde{f}} \quad , \quad
    \tilde{a}=\sqrt{2}\tilde{f} \quad ,
\end{equation}
\begin{equation}\label{}
    \widetilde{N}_{4,5}=\frac{2\sqrt{2}}{\sqrt{32\tilde{f}^2
                        +\lambda\Big(\lambda\mp\sqrt{32\tilde{f}^2+\lambda^2}\Big)}} \quad , \quad
    \widetilde{X}_{4,5}=-\frac{1}{4}\Big(\lambda\mp\sqrt{32\tilde{f}^2+\lambda^2}\Big) \quad ,
\end{equation}
where $\lambda$ and $\tilde{f}$ represent parameters of the Higgs potential~\cite{ScalarS}, which we have assumed of the same order: $\lambda \equiv \lambda_{9}=\lambda_{17}=\lambda_{19} \sim O(1)$ and $\tilde{f}\equiv\tilde{f}_1=\tilde{f}_2\sim O(1)$.

\begin{table}[!ht]
\centering
\caption{\label{Table1} Values of $Q_S$ and $g^S_{V^0}$ in the couplings of $Z$ and $Z'$ to charged scalars.}
\begin{tabular}{ccccc}\hline\hline
$S$           & $Q_S$ & $g_Z^S$     & $g_{Z'}^S$                \\ \hline
$G_W^+$       & 1     & $c_{2W}$    & $-\frac{c_Wc_\theta}{\sqrt{3}}t_{\theta}^2$   \\
$G_Y^+$       & 1     & $-1-2s_W^2$ & $\frac{c_Wc_\theta}{\sqrt{3}}(1-2t_\theta^2)$ \\
$G_Y^{++}$    & 2     & $1-4s_W^2$  & $\frac{c_Wc_\theta}{\sqrt{3}}(1-2t_\theta^2)$ \\
$h_1^+,h_4^+$ & 1     & $-2s_W^2$   & $\frac{2c_Wc_\theta}{\sqrt{3}}$ \\
$h_2^+$       & 1     & $c_{2W}$    & $\frac{c_Wc_\theta}{\sqrt{3}}[1-2(1+t_\theta^2)H_{22}^2]$ \\
$h_3^+$       & 1     & $c_{2W}$    & $\frac{c_Wc_\theta}{\sqrt{3}}[1-2(1+t_\theta^2)H_{32}^2]$ \\
$d_1^{++}$    & 2     & $-4s_W^2$   & $\frac{2c_Wc_\theta}{\sqrt{3}}\widetilde{N}_4^2[2\widetilde{X}_4^2+(1-t_\theta^2)\tilde{a}^2]$ \\
$d_2^{++}$    & 2     & $-4s_W^2$   & $\frac{2c_Wc_\theta}{\sqrt{3}}\widetilde{N}_5^2[2\widetilde{X}_5^2+(1-t_\theta^2)\tilde{a}^2]$ \\
$d_3^{++}$    & 2     & $2c_{2W}$   & $\frac{2c_Wc_\theta}{\sqrt{3}}$ \\ \hline\hline
\end{tabular}
\end{table}

\section{The amplitude for the $V^0\to \gamma \gamma \gamma$ decay}
\label{a} 
\begin{figure}
\centering
\includegraphics[width=4cm]{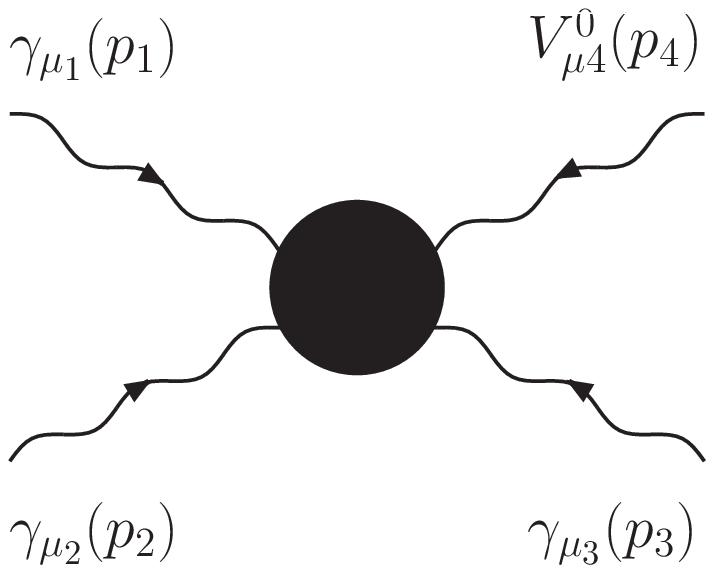}
\caption{\label{F1} Notation and conventions for the $V^0\gamma \gamma\gamma$ vertex.}
\end{figure}

\begin{figure}
\centering
\includegraphics[width=6cm]{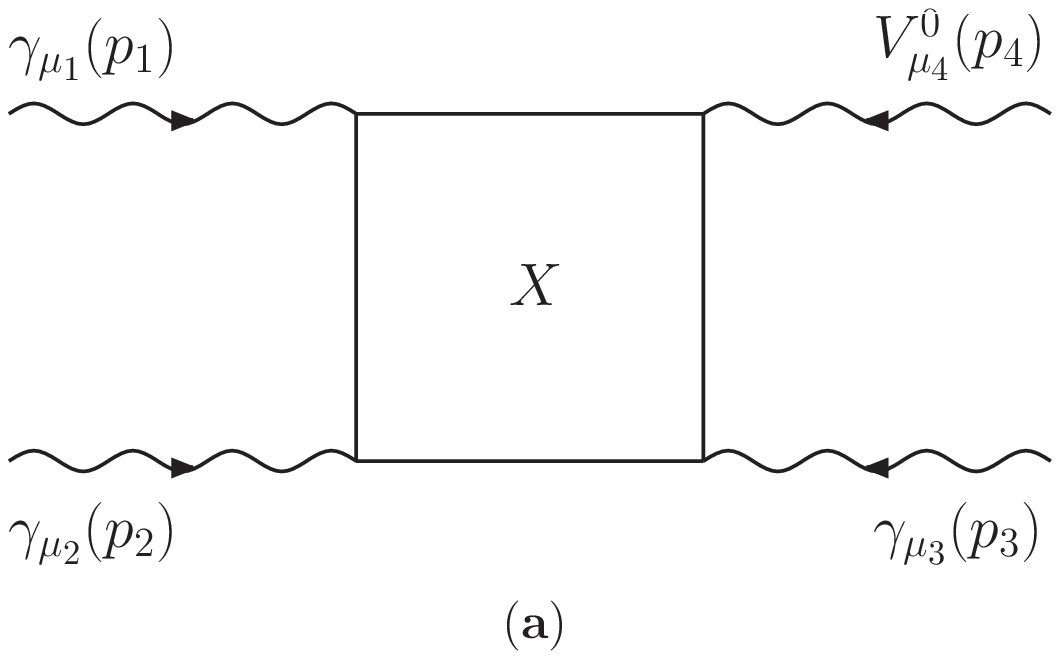} \ \ \
\includegraphics[width=6cm]{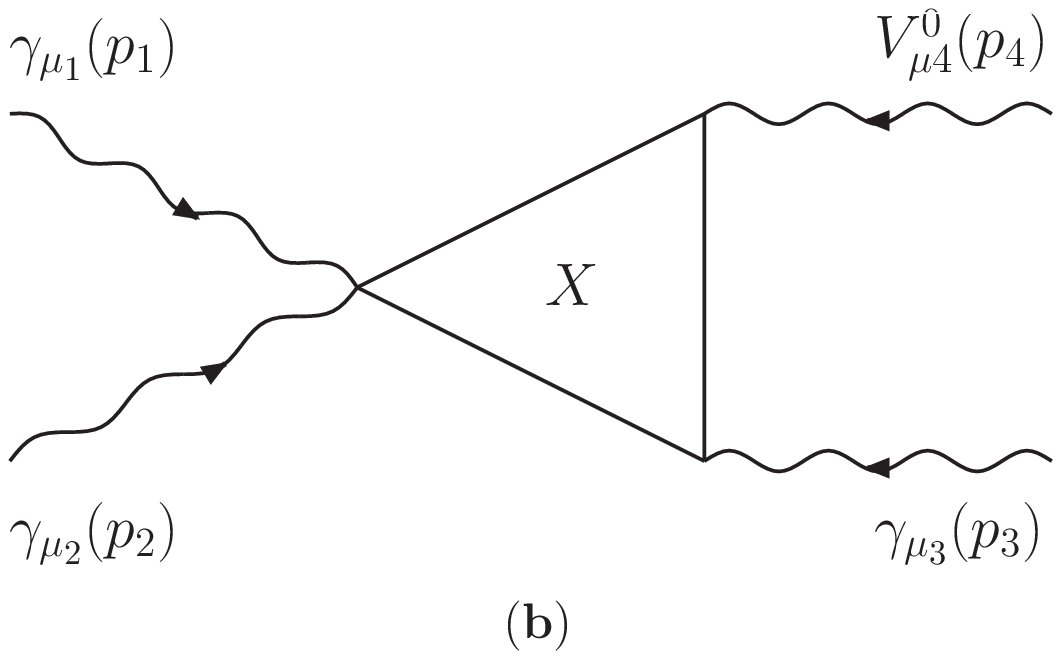}
\includegraphics[width=6cm]{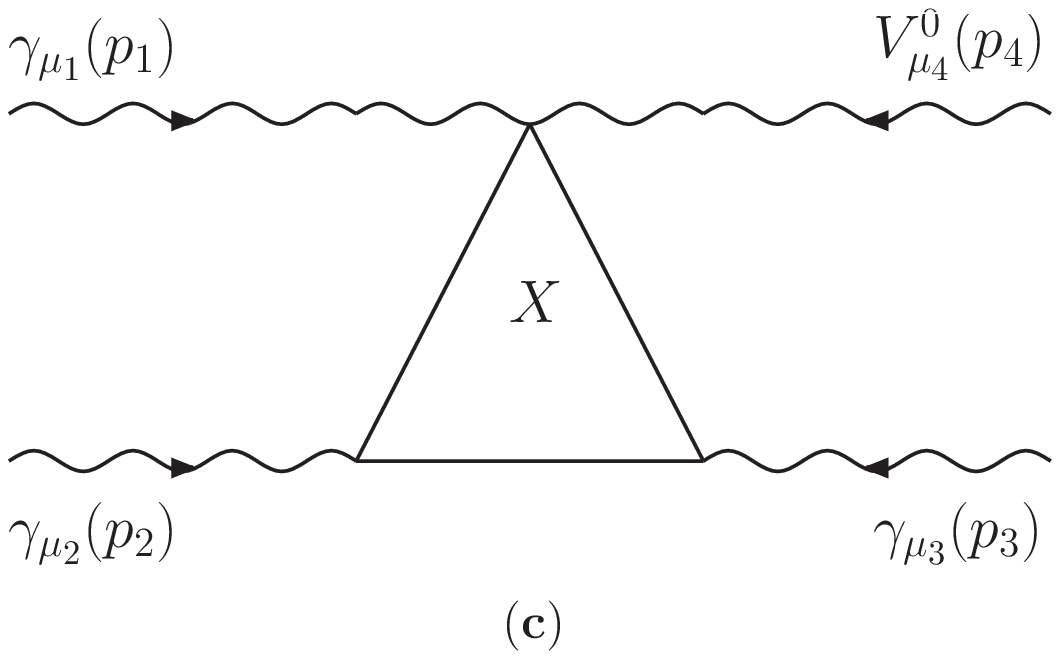} \ \ \
\includegraphics[width=6cm]{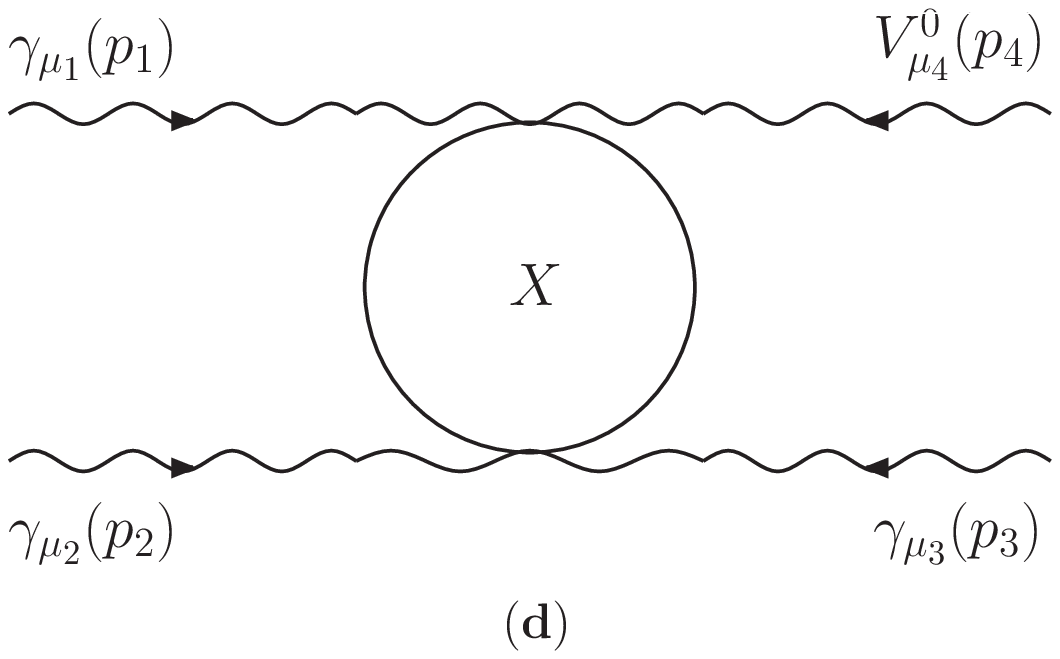}
\caption{\label{F2}Generic Feynman diagrams contributing to the $V^0 \to \gamma \gamma \gamma$ decay. Fermions only contribute through box diagrams as shown in $(a)$.}
\end{figure}

In this section, we present the amplitudes for the on-shell vertices $V^0\gamma \gamma \gamma$, with $V^0=Z,Z'$. Our notation and conventions are established in Fig. \ref{F1}. These couplings first arise at the one-loop level through diagrams shown in Fig. \ref{F2}, in which circulate all charged particles of the model. The invariant amplitude can be written as follows:

\begin{equation}
\mathcal{M}_{\gamma\gamma\gamma V^0}=\sum_X\mathcal{M}^{\mu_1\mu_2\mu_3\mu_4}_{\gamma\gamma\gamma V^0}
\epsilon_{\mu_1}(p_1,\lambda_1)\epsilon_{\mu_2}(p_2,\lambda_2)
\epsilon_{\mu_3}(p_3,\lambda_3)\epsilon_{\mu_4}(p_4,\lambda_4) \ ,
\end{equation}
where $X$ denotes the type of particles circulating in the loops. The fermions will be collectively denoted by $F=u$, $d$, $s$, $c$, $b$, $t$, $D$, $S$, $T$, $e$, $\mu$, $\tau$. On the other hand, gauge bosons and their associated pseudo Goldstone bosons and ghosts contributions will be separated into vector and scalar contributions, namely, vector ($V$) and scalar ($S$) particles, with $V=W^+$, $Y^+$, $Y^{++}$, and $S=G_W^+$, $C_W^+$, $\bar{C}_W^+$, $G_Y^+$, $C_Y^+$, $\bar{C}_Y^+$, $G_Y^{++}$, $C_Y^{++}$, $\bar{C}_Y^{++}$. The contribution of the physical Higgs bosons will be denoted by $H=h_1^+$, $h_2^+$, $h_3^+$, $h_4^+$, $d_1^{++}$, $d_2^{++}$, $d_3^{++}$. For calculation purposes, it is convenient to organize the amplitudes according to the spin of particles circulating in the loops. This is possible since we calculate the gauge particles contribution using covariant gauges $R_\xi$-gauges~\cite{T1,HT,T6,T7}, which separately render finite and gauge invariant the pseudo Goldstone bosons and ghosts contributions
\begin{equation}
\mathcal{M}_{\gamma\gamma\gamma V^0}=\mathcal{M}_{\frac{1}{2}}+\mathcal{M}_1+\mathcal{M}_0 \ ,
\end{equation}
where $\mathcal{M}_{\frac{1}{2}}$, $\mathcal{M}_1$, and $\mathcal{M}_0$ are, respectively, the spinorial, vectorial, and scalar amplitudes, which can be written as follows:
\begin{equation}
\mathcal{M}_{\frac{1}{2}}=\sum_F\mathcal{M}_F \ ,
\end{equation}
\begin{equation}
\mathcal{M}_1=\sum_V\mathcal{M}_V \ ,
\end{equation}
\begin{equation}
\mathcal{M}_0=\sum_{S,H}\mathcal{M}_{S,H} \ .
\end{equation}
Structurally speaking, these are the only amplitudes that are different. The Lorentz tensor structure of the amplitudes, as well as the functional way of the form factors involved, are dictated by electromagnetic gauge invariance and Bose symmetry. Gauge invariance requires that
\begin{equation}
p_{i\, \mu_i}\sum_X\mathcal{M}^{\mu_1\mu_2\mu_3\mu_4}_{\gamma\gamma\gamma V^0}=0\ ,\ i=1,2,3\ ,
\end{equation}
whereas Bose statistics dictates that the amplitudes must be symmetric under the interchanges of pairs of photons:
\begin{equation}
(p_1,\mu_1)\leftrightarrow (p_2,\mu_2)\leftrightarrow (p_3,\mu_3) \quad .
\end{equation}
There are six different configurations by each type of diagram.

The fermionic contribution is given by six box diagrams as that shown in Fig. \ref{F2}$(a)$. The corresponding tensorial amplitude is given by
\begin{equation}\label{}
    \mathcal{M}_{\frac{1}{2}}^{\mu_1\mu_2\mu_3\mu_4}=\sum_F \mathcal{M}_{F,Box}^{\mu_1\mu_2\mu_3\mu_4} \ .
\end{equation}

In the case of the vector contribution all types of diagrams in Fig. \ref{F2} contribute. The tensorial amplitude can be written as
\begin{equation}\label{}
    \mathcal{M}_1^{\mu_1\mu_2\mu_3\mu_4}=\sum_V \bigg( \mathcal{M}_{V,Box}^{\mu_1\mu_2\mu_3\mu_4}
        +\mathcal{M}_{V,Triangle1}^{\mu_1\mu_2\mu_3\mu_4}+\mathcal{M}_{V,Triangle2}^{\mu_1\mu_2\mu_3\mu_4}+
        \mathcal{M}_{V,Bubble}^{\mu_1\mu_2\mu_3\mu_4} \bigg) \ .
\end{equation}
The Feynman rules needed for the vector contribution are given in Refs.~\cite{T1,HT,T6,T7}, these and other new rules are summarized at the Appendix~\ref{FR}.

Finally, the scalar contributions are characterized by the following amplitudes
\begin{equation}\label{}
    \mathcal{M}_0^{\mu_1\mu_2\mu_3\mu_4}=\sum_X \bigg( \mathcal{M}_{X,Box}^{\mu_1\mu_2\mu_3\mu_4}
        +\mathcal{M}_{X,Triangle1}^{\mu_1\mu_2\mu_3\mu_4}+\mathcal{M}_{X,Triangle2}^{\mu_1\mu_2\mu_3\mu_4}+
        \mathcal{M}_{X,Bubble}^{\mu_1\mu_2\mu_3\mu_4} \bigg) \ ,
\end{equation}
here $X$ stands for a nonphysical scalar $S$ or a Higgs boson $H$. The diverse vertices involved in the above amplitudes were given in the previous section for the case of the couplings of $V^0$ with scalars, whereas the corresponding couplings of the photon are dictated by scalar electrodynamics.

Once the loop integrals are solved, the amplitudes can be expressed in terms of gauge structures and their associated form factors as follows:
\begin{equation}\label{AmplitudTensorialCompleta}
\mathcal{M}_{\gamma\gamma\gamma V^0}^{\mu_1\mu_2\mu_3\mu_4}=\frac{i}{\pi^2}\frac{ge^3}{2 c_W}\sum_{i=1}^{18}
                                       F_{V^0i} T_i^{\mu_1\mu_2\mu_3\mu_4} \ ,
\end{equation}
where
\begin{equation}
F_{V^0i} = F_{V^0i}^{\frac{1}{2}}+F_{V^0i}^1+F_{V^0i}^0 \ .
\end{equation}
The Lorentz tensors $T_i^{\mu_1\mu_2\mu_3\mu_4}$ are gauge structures given by~\cite{T4}
\begin{eqnarray}
T^{\mu_1\mu_2\mu_3\mu_4}_{1}&=&(p_1\cdot p_2 g^{\mu_1\mu_2}-p_2^{\mu_1} p_1^{\mu_2}) (p_1\cdot p_3 g^{\mu_3\mu_4}-p_1^{\mu_3} p_3^{\mu_4}) \ , \\
T^{\mu_1\mu_2\mu_3\mu_4}_{7}&=&(p_1\cdot p_3 p_2^{\mu_1}-p_1\cdot p_2 p_3^{\mu_1}) (p_2\cdot p_3 g^{\mu_2\mu_3}-p_3^{\mu_2} p_2^{\mu_3}) p_2^{\mu_4} \ , \\
T^{\mu_1\mu_2\mu_3\mu_4}_{13}&=&(p_1\cdot p_3 g^{\mu_1\mu_2}-p_3^{\mu_1}p_1^{\mu_2})(p_2\cdot p_3g^{\mu_3\mu_4}-p_2^{\mu_3}p_3^{\mu_4})
\nonumber\\
&& +(p_1\cdot p_2 p_3^{\mu_1}-p_1\cdot p_3p_2^{\mu_1})(p_3^{\mu_2}g^{\mu_3\mu_4}-p_3^{\mu_4}g^{\mu_2\mu_3}) \ .
\end{eqnarray}
Using Bose symmetry, each of these tensors determines a set of 6 gauge structures and their associated form factors:
\begin{eqnarray}
&&\{F_{V^01}T_{1}^{\mu_1\mu_2\mu_3\mu_4},...,F_{V^0 6}T_{6}^{\mu_1\mu_2\mu_3\mu_4}\}\; , \nonumber \\
&&\{F_{V^07}T_{7}^{\mu_1\mu_2\mu_3\mu_4},...,F_{V^0 12}T_{12}^{\mu_1\mu_2\mu_3\mu_4}\}\; , \nonumber \\
&&\{F_{V^013}T_{13}^{\mu_1\mu_2\mu_3\mu_4},...,F_{V^0 18}T_{18}^{\mu_1\mu_2\mu_3\mu_4}\}\;.
\end{eqnarray}
As it can easily be verified, all these Lorentz tensors are gauge structures, as they satisfy the transversality condition:
\begin{equation}\label{}
    p_{i\mu_i}T^{\mu_1\mu_2\mu_3\mu_4}=0 \ , \qquad i=1, 2, 3.
\end{equation}
Notice that the amplitude (\ref{AmplitudTensorialCompleta}) is also transverse with respect to the $V^0$ gauge boson, which means that it appears in the $V^0\gamma \gamma \gamma$ coupling not directly, but only through the strength tensor $V^0_{\mu \nu}=\partial_\mu V^0_\nu-\partial_\nu V^0_\mu$.

The fermionic contribution is given by
\begin{equation}
\mathcal{M}_{\frac{1}{2}}^{\mu_1\mu_2\mu_3\mu_4}=\frac{i}{\pi^2}\frac{ge^3}{2 c_W}\sum_{i=1}^{18}
                                       F_{V^0i}^{\frac{1}{2}} T_i^{\mu_1\mu_2\mu_3\mu_4} \ ,
\end{equation}
where
\begin{eqnarray}\label{FermionicFormFactor}
F_{V^0i}^{\frac{1}{2}} &=& \sum_{F=q, Q, l} g_{\frac{1}{2}}^F f_{V^0i}^F \nonumber \\
                     &=& F_{V^0i}^q+F_{V^0i}^Q+F_{V^0i}^l \, ,
\end{eqnarray}
with the definition
\begin{equation}\label{FermionicConstant}
g_{\frac{1}{2}}^F\equiv -N_F Q_F^3 g_{V^0}^F \ ,
\end{equation}
being $N_F$ the color number, $1$ for leptons and $3$ for quarks, $Q_F$ is the charge content in units of $e$, and $g_{V^0}^F$ is the vector-like coupling of $V^0$ to fermions, which is given in~\cite{T2}.
In (\ref{FermionicFormFactor}) $q$, $Q$, and $l$ stand for known quarks, exotic quarks, and charge leptons, respectively. The form factors $f_{V^0i}^F$ are exactly the ones given in~\cite{T4} and we refrain from present here.

As far as the gauge boson contributions is concerned, we first note that in nonlinear gauges the contributions of pseudo Goldstone bosons, ghosts and antighosts satisfy the following relations
\begin{equation}\label{}
    \mathcal{M}_{G}=-\mathcal{M}_{C}=-\mathcal{M}_{\bar{C}} \ .
\end{equation}
Then, the spin 1 contribution comprises the vectorial amplitudes plus the nonphysical scalar ones:
\begin{equation}
\mathcal{M}_1=\sum_{V,S}\mathcal{M}_{V,S} \ .
\end{equation}
Once the corresponding vectorial and scalar amplitudes are collected, the tensorial amplitude can be written as:
\begin{equation}
\mathcal{M}_{1}^{\mu_1\mu_2\mu_3\mu_4}=\frac{i}{\pi^2}\frac{ge^3}{2 c_W}\sum_{i=1}^{18}
                                       F_{V^0i}^{1} T_i^{\mu_1\mu_2\mu_3\mu_4} \ ,
\end{equation}
where
\begin{eqnarray}
F_{V^0i}^1 &=& \sum_{V=W^+,Y^+,Y^{++}} g_1^V f_{V^0i}^V \
               -\sum_{S=G_W^+,G_Y^+,G_Y^{++}} g_0^S f_{V^0i}^S \nonumber\\
         &=& F_{V^0i}^W+F_{V^0i}^Y \ ,
\end{eqnarray}
with
\begin{equation}\label{BosonicConstant}
    g_1^V\equiv Q_V^3 g_{V^0}^V \quad , \quad g_0^S\equiv Q_S^3 g_{V^0}^S \ .
\end{equation}

More explicitly,
\begin{equation}
F_{V^0i}^W = Q_{W^+}^3\big(g_{V^0}^{W^+} f_{V^0i}^{W}-g_{V^0}^{G_W^+} f_{V^0i}^{G_W}\big) \ ,
\end{equation}

\begin{equation}
F_{V^0i}^Y = \Big(Q_{Y^+}^3 g_{V^0}^{Y^+}+Q_{Y^{++}}^3 g_{V^0}^{Y^{++}}\Big)f_{V^0i}^Y
           -\Big(Q_{Y^+}^3 g_{V^0}^{G_Y^+}+Q_{Y^{++}}^3 g_{V^0}^{G_Y^{++}}\Big)f_{V^0i}^{G_Y} \ ,
\end{equation}
and $m_Y\equiv m_{Y^+}=m_{Y^{++}}$. Particularly,

\begin{eqnarray}
F_{Zi}^W &=& 2c_W^2f_{Zi}^{W}-c_{2W}f_{Zi}^{G_W} \ ,
\end{eqnarray}
\begin{eqnarray}
F_{Z'i}^W &=& 0 \ ,
\end{eqnarray}
\begin{eqnarray}
F_{Zi}^Y &=& (7-34s_W^2)\big(f_{Zi}^Y-f_{Zi}^{G_Y}\big) \ ,
\end{eqnarray}
\begin{eqnarray}
F_{Z'i}^Y &=& 9\sqrt{3}\sqrt{1-4s_W^2}f_{Z'i}^Y-3\sqrt{3}c_Wf_{Z'i}^{G_Y} \ .
\end{eqnarray}
The various form factors appearing in the above expressions are given in the Appendix~\ref{FFactors}.

On the other hand, the contribution of the charged Higgs bosons is given by

\begin{eqnarray}
F_{V^0i}^0 &=& \sum_{H=h_1^+,h_2^+,h_3^+,h_4^+,d_1^{++},d_2^{++},d_3^{++}} g_0^H f_{V^0i}^H \nonumber\\
         &=& F_{V^0i}^H \ ,
\end{eqnarray}
with
\begin{equation}\label{HiggsConstant}
    g_0^H\equiv Q_H^3 g_{V^0}^H \ ,
\end{equation}
where
\begin{equation}
F_{Zi}^H = 2(9-52s_W^2)f_{Zi}^h \ ,
\end{equation}

\begin{equation}
F_{Z'i}^H = \frac{69c_W}{\sqrt{3}} f_{Z'i}^h \ ,
\end{equation}
and $m_H\equiv m_{h^+}=m_{d^{++}}$. The form factors $f_{Zi}^h$ and $f_{Z'i}^h$ are given in the Appendix~\ref{FFactors}.

The differential decay width is given by
\begin{equation}\label{}
    \frac{d^2\Gamma}{dxdy}=\frac{m_{V^0}}{256\pi^3}|\mathcal{M}|^2 \ ,
\end{equation}
 where $x\equiv 2p_1^0/m_{V^0}$, $y\equiv 2p_2^0/m_{V^0}$, $z\equiv 2 p_3^0/m_{V^0}$, with $x+y+z=2$. This parametrization leads to
$p_1\cdot p_2=m_{V^0}^2(x+y-1)/2$ \ , \
$p_1\cdot p_3=m_{V^0}^2(1-y)$/2 \ , \
$p_2\cdot p_3=m_{V^0}^2(1-x)/2$ \ ,
so the allowed region is determined by the limits $0\leq x\leq 1$ \ and \ $1-x \leq y \leq 1$. This leads to a decay width given by
\begin{eqnarray}
\Gamma(V^0\to\gamma\gamma\gamma) &=& \frac{m_{V^0}}{256\pi^3}\frac{1}{3!}\int_0^1\int_{1-x}^1
|\mathcal{M}_{V^0\to\gamma\gamma\gamma}|^2dydx \ ,
\end{eqnarray}
where $1/3!$ is a symmetry factor due to the presence of three identical particles in the final state. In the above expression,
\begin{equation}
\mathcal{M}_{V^0\to\gamma\gamma\gamma}=i\frac{8\alpha^2}{c_W s_W}\sum_X \mathcal{V}_X \ ,
\end{equation}
where
\begin{equation}\label{}
\mathcal{V}_X\equiv \sum_{i=1}^{18}F_{V^0i}^X T_{V^0i}^{\mu_1\mu_2\mu_3\mu_4}
\epsilon_{\mu_1}^*(p_1,\lambda_1)\epsilon_{\mu_2}^*(p_2,\lambda_2)\epsilon_{\mu_3}^*(p_1,\lambda_3)
\epsilon_{\mu_4}(p_1,\lambda_4) \ ,
\end{equation}
with $X$ denoting the virtual particles contributing to the $V^0\to \gamma \gamma \gamma$ transition. The squared of the amplitude takes the form
\begin{eqnarray}
|\mathcal{M}_{V^0\to\gamma\gamma\gamma}|^2 &=& \bigg(\frac{8\alpha^2}{c_Ws_W}\bigg)^2\Bigg[ \sum_X|\mathcal{V}_X|^2
+\sum_{X\neq X'} 2\mathrm{Re}\big( \mathcal{V}_X\mathcal{V}_{X'}^* \big) \Bigg] \ ,
\end{eqnarray}
where
\begin{eqnarray}
\mathcal{V}_X\mathcal{V}_{X'}^* &=& \frac{1}{3}\sum_{\lambda_1,\lambda_2,\lambda_3,\lambda_4}\sum_{i,j=1}^{18}
F_{V^0i}^XF_{V^0j}^{X'*}T_{V^0i}^{\mu_1\mu_2\mu_3\mu_4}T_{V^0j}^{\nu_1\nu_2\nu_3\nu_4}
\epsilon_{\mu_1}^*(p_1,\lambda_1)\epsilon_{\nu_1}(p_1,\lambda_1')
\epsilon_{\mu_2}^*(p_2,\lambda_2)\epsilon_{\nu_2}(p_2,\lambda_2') \nonumber\\
&&\quad \times \epsilon_{\mu_3}^*(p_3,\lambda_3)\epsilon_{\nu_3}(p_3,\lambda_3')
\epsilon_{\mu_4}(p_4,\lambda_4)\epsilon_{\nu_4}^*(p_4,\lambda_4') \ ,
\end{eqnarray}
where the factor $1/3$ results from averaging on the polarization states of $V^0$. In terms of the $\mathcal{V}_X$ amplitudes, the decay width can be written as follows:
\begin{eqnarray}
\Gamma(V^0\to\gamma\gamma\gamma) &=& \frac{m_{V^0}}{256\pi^3}\frac{1}{3!}\bigg(\frac{8\alpha^2}{c_Ws_W}\bigg)^2 \int_0^1\int_{1-x}^1 \Bigg[ \sum_X|\mathcal{V}_X|^2
+\sum_{X\neq X'} 2\mathrm{Re}\big( \mathcal{V}_X\mathcal{V}_{X'}^* \big) \Bigg] dydx \ ,
\end{eqnarray}

\section{Results and discussion}
\label {rd} We now turn to discuss our results. We first discuss the impact of new physics on the $Z\to \gamma \gamma \gamma$ decay. We have verified that any virtual contribution $X$ such that $m_X>m_Z/2$ is marginal. To estimate these new physics effects, we consider a scenario with $m_Q$=500 GeV and $m_Y$=$m_H$=250 GeV. The relative importance of each type of contribution is shown in Tables \ref{TABLEPS} and \ref{TABLEFB}. Our results for the SM contributions is in perfect agreement with those given in the literature~\cite{Z3gFermionA,Z3gFermionWE,Z3gW}. From these tables, it can be appreciated that the contributions of the $u$ and $c$ quarks dominate. The contributions of the $W$ gauge boson, as well as those arising from the charged leptons and the $d$, $s$ and $b$ quarks are all of the same order of magnitude and one order of magnitude lower than those given by the $u$ and $c$ quarks. It can be seen that the top quark contribution is marginal indeed, as it is two orders of magnitude lower than those induced by the $u$ and $c$ quarks. As far as new physics effects are concerned, the effect on this decay is quite suppressed, as their contribution is three orders of magnitude lower than the SM contribution, at the best.

\begin{table}
\caption{\label{TABLEPS} Contributions to the $\mathrm{Br}(Z\to \gamma\gamma\gamma)$ in the scenario $m_{Q}=500$ GeV, $m_Y=m_H=250$ GeV. Interference effects among sectors are also shown.}
\begin{ruledtabular}
\begin{center}
 \begin{tabular}{|l|c|}
  Sector & Br \\
  \hline
  Fermions   & $4.16\times 10^{-10}$ \\
  \hline
  Gauge Bosons   & $1.03\times 10^{-11}$ \\
  \hline
  Scalar  & $3.04\times 10^{-13}$ \\
  \hline
  Fermions-Gauge Bosons & $9.92\times10^{-11}$ \\
  \hline
  Fermions-Scalar & $-5.90\times 10^{-14}$ \\
  \hline
  Gauge Bosons-Scalar & $4.30\times 10^{-13}$ \\
  \hline
  Total & $5.26\times 10^{-10}$
\end{tabular}
\end{center}
\end{ruledtabular}
\end{table}

\begin{table}
\caption{\label{TABLEFB} Fermionic and bosonic contributions to the $\mathrm{Br}(Z\to \gamma\gamma\gamma)$ in the minimal 331 model. Some interference effects are also shown.}
\begin{ruledtabular}
\begin{center}
 \begin{tabular}{|l|c|c|c|c|c|c|c|}
  Fermions & Br & Quarks & Br & SM Quarks & Br & Bosons & Br  \\
  \hline
  Quarks   & $2.67\times 10^{-10}$ & SM quarks & $2.67\times 10^{-10}$ & u & $3.76\times 10^{-11}$ & W boson & $1.03\times10^{-11}$ \\
  \hline
  Leptons   & $1.66\times 10^{-11}$ & Exotic quarks & $6.70\times 10^{-14}$ & c & $4.08\times 10^{-11}$ & Bileptons & $1.88\times10^{-13}$ \\
  \hline
  Quarks-Leptons   & $1.33\times 10^{-10}$ & SM-Exotic quarks & $-3.62\times 10^{-14}$ & t & $1.11\times 10^{-13}$ & W boson-Bileptons & $-1.57\times10^{-13}$ \\
  \hline
   &  &    &  & d & $1.91\times10^{-12}$ &  &  \\
  \hline
   &  &    &  & s & $1.91\times10^{-12}$ &  & \\
  \hline
   &  &  &  & b & $2.12\times10^{-12}$ &  &  \\
  \hline
   &  &  &  & Interference & $1.82\times10^{-10}$ &  &  \\
  \hline
  $\mathrm{Total}$ & $4.16\times 10^{-10}$ &  & $2.67\times 10^{-10}$ &  & $2.67\times10^{-10}$ &  & $1.03\times10^{-11}$
\end{tabular}
\end{center}
\end{ruledtabular}
\end{table}

\begin{figure}[!ht]
\begin{center}
\includegraphics[width=8.2cm]{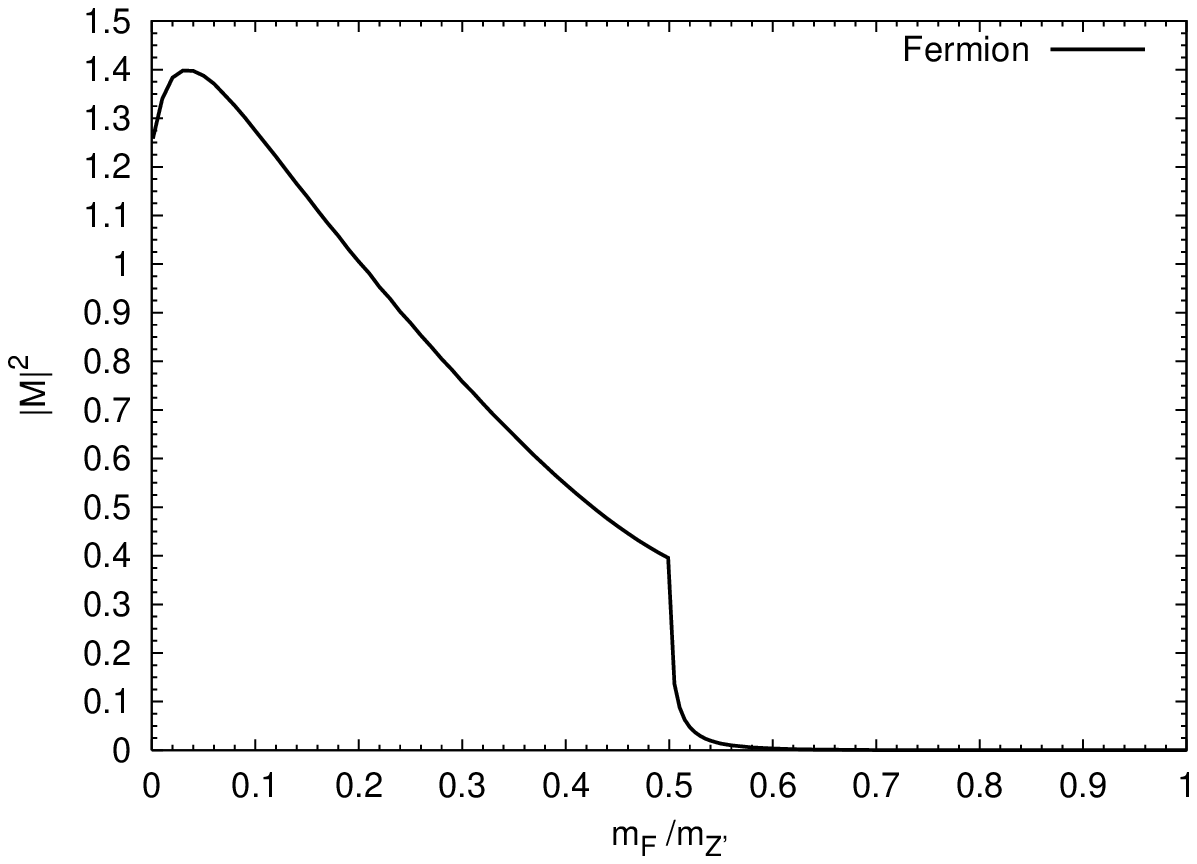}
\includegraphics[width=8.2cm]{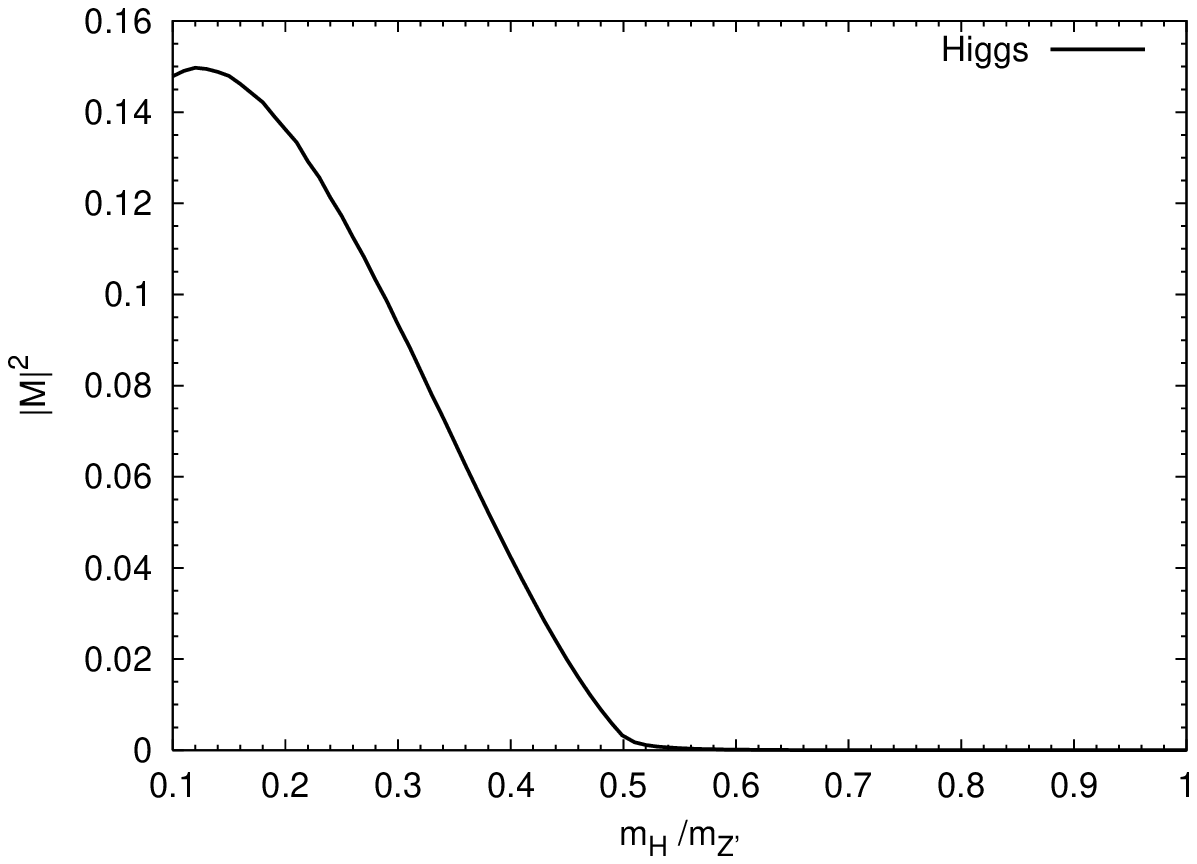}
\includegraphics[width=8.2cm]{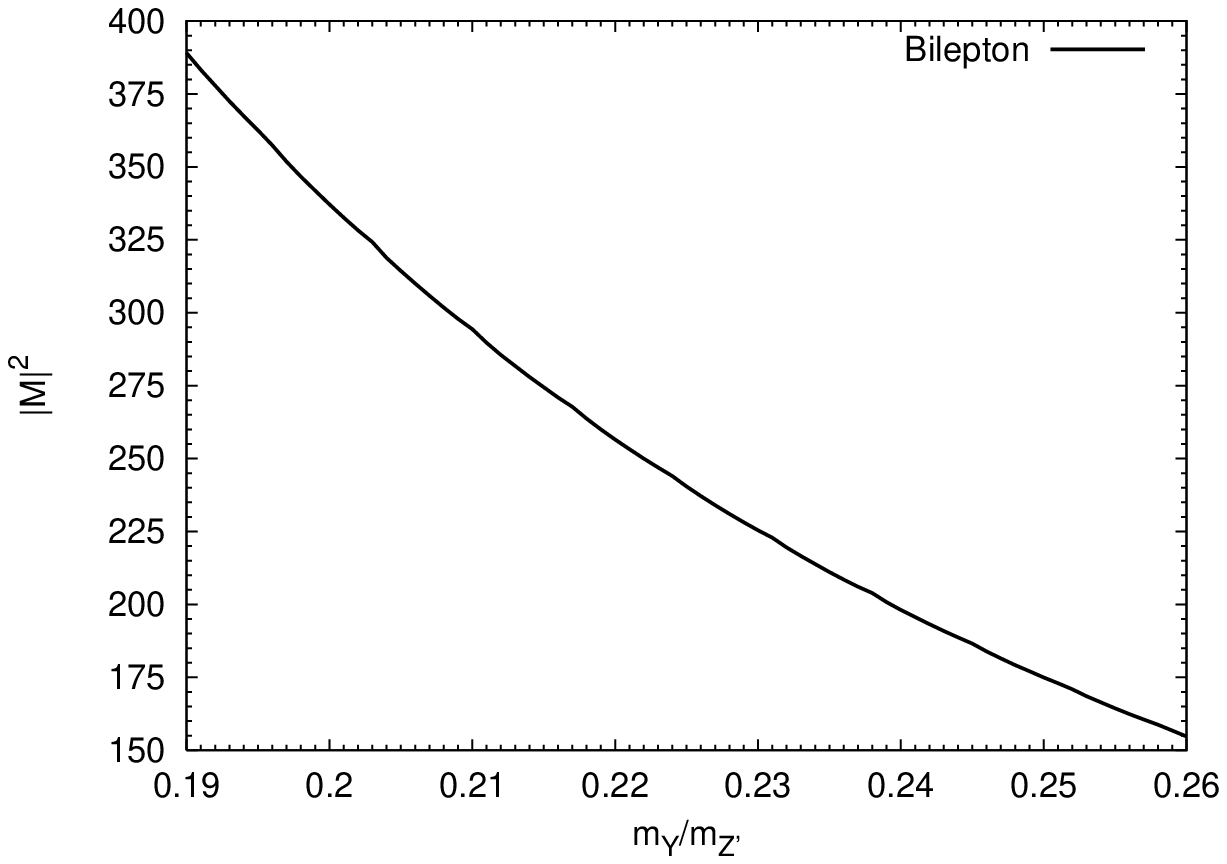}
\caption{\label{F3}Sensitivity of the $Z'\to \gamma \gamma \gamma$ decay to the spin of the particle circulating in the loop. The combined effects of spin and mass ratio $m/m_{Z'}$ are shown: first graphic, spin $1/2$ effect, second graphic, spin $0$ effect, and third graphic, spin $1$ effect.}
\end{center}
\end{figure}

We now turn to discuss our results for the $Z'\to \gamma \gamma \gamma$ decay. Before analyzing the diverse type of contributions to the branching fraction of this decay, it is interesting  to examine the role played by the spin of the particles circulating in the loop. As it was discussed in the previous section, the diverse contributions to the $Z'\to \gamma \gamma \gamma$ decay can be grouped in accordance with the spin of the particles circulating in the loop, namely, spin $0$ (charged scalars), spin $1/2$ (charged leptons and quarks), and spin $1$ (simple and doubly charged bileptons). Then, we analyze the behavior of the squared amplitude for only one type of particle as a function of the mass ratio $m/m_{Z'}$, with $m$ the mass of the particle circulating in the loop. The squared amplitude which we will examine is given by
\begin{eqnarray}
|\mathrm{M}|^2 &\equiv& \int_0^1\int_{1-x}^1
\big|\mathrm{M}_{Z'\to\gamma\gamma\gamma}\big|^2dydx \ ,
\end{eqnarray}
where
\begin{equation}
\mathrm{M}_{Z'\to\gamma\gamma\gamma}^{\mu_1\mu_2\mu_3\mu_4} = \sum_{i=1}^{18}f_{Z'i}^\mathrm{Spin}
    T_i^{\mu_1\mu_2\mu_3\mu_4} \, .
\end{equation}
In this expression, $f_{Z'i}^\mathrm{Spin}$ are numerical factors, which are irrelevant for the present discussion. In Fig. \ref{F3}, the behavior of $|\mathrm{M}|^2$ as a function of $m/m_{Z'}$ for the three types of spins is shown. From these figures, it can be appreciated that in the case of spin $1/2$ the highest contribution occurs for $m_f/m_{Z'}=0.03$, which is about of $1.4$. This means that the mean contribution from the fermionic sector to the $Z'\to \gamma \gamma \gamma$ decay arises from the lightest SM charged leptons and quarks. On the other hand, from the second graphic of the same figure, it can be seen that the highest spin $0$ effect ($0.15$) occurs for  $m_H/m_{Z'}=0.12$. This shows that the main contribution would arise from a relatively light charged Higgs boson, with a mass of about a $12\%$ of the $m_{Z'}$ mass. It is interesting to see that the scalar contribution is about of one order of magnitude lower than the fermionic one. As far as the spin $1$ contribution is concerned, it deserves special attention. In first place is the fact that the $Z'$ particle does not couple directly to pairs of SM $W$ gauge bosons at the tree level, but only very weakly through the $Z'-Z$ mixing. The absence of a direct coupling $Z'WW$ is a consequence of the fact that the $Z'$ gauge boson emerges in the first stage of spontaneous symmetry breaking as a singlet of $SU(2)_L$. So, the only spin $1$ contributions to the $Z'\to \gamma \gamma \gamma$ decay arise from the simple and doubly charged bileptons. On this matter, one interesting feature of the minimal 331 model is that the new gauge boson masses are bounded from above due to the theoretical constraint which yields $s^2_W\leqslant 1/4$~\cite{F331,Ng1}. The fact that the value of $s^2_W$ is very close to $1/4$ at the $m_{Z'}$ scale leads to an upper bound on the scale associated with the first stage of SSB, which translates directly into a bound on the $Z'$ mass given by $m_{Z'}\leqslant 3.1$ TeV~\cite{Ng1}, which in turns implies that the bilepton masses cannot be heavier than $m_{Z'}/2$~\cite{Ng1}. It turns out to be that this peculiar structure of the model imposes the theoretical restriction $m_Y/m_{Z'}<0.26$~\cite{FLRN}, whereas lower bounds on $m_Y$ and $m_{Z'}$ obtained from experimental data restrict this mass ratio to be $0.19<m_Y/m_{Z'}$~\cite{FLRN}. The behavior of the spin $1$ amplitude as a function of the $m_Y/m_{Z'}$ is shown in the third graph of Fig. \ref{F3} within the allowed interval $0.19<m_Y/m_{Z'}<0.26$. From this graph, it can be appreciated that $|M|^2$ ranges from $389$ to $155$ within this interval. At this level of amplitude, the vector contribution is larger by more than two and three orders of magnitude than the fermionic and scalar contributions, respectively. As we will see below, this dominant effect of the vector particle is reinforced by the exotic charge contained of one of the bileptons.

\begin{figure}[!ht]
\begin{center}
\includegraphics[width=8cm]{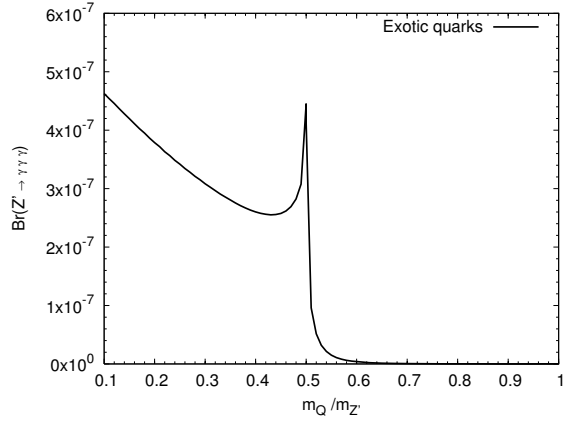}
\caption{\label{F4}Exotic quark contribution to $Br(Z'\to \gamma \gamma \gamma)$ as a function of $m_Q/m_{Z'}$.}
\end{center}
\end{figure}

\begin{figure}[!ht]
\begin{center}
\includegraphics[width=8cm]{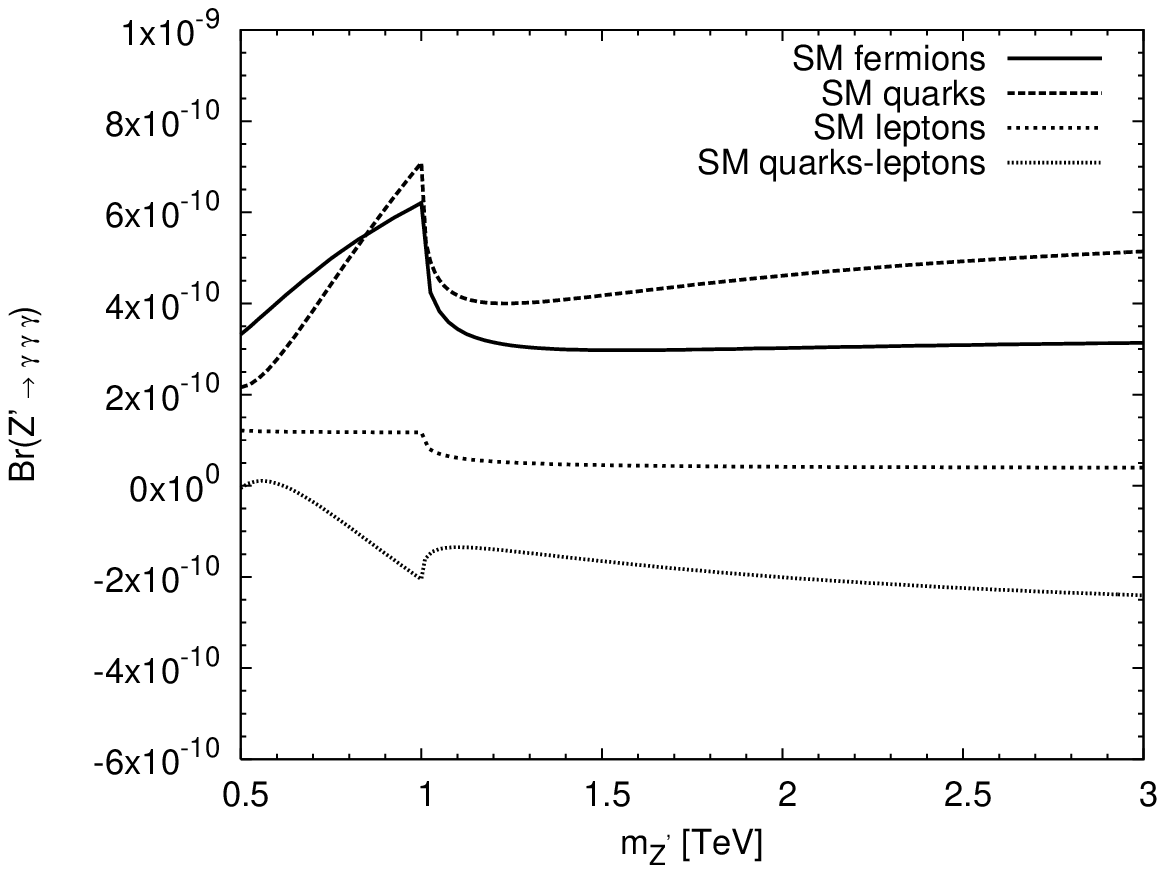}
\includegraphics[width=8cm]{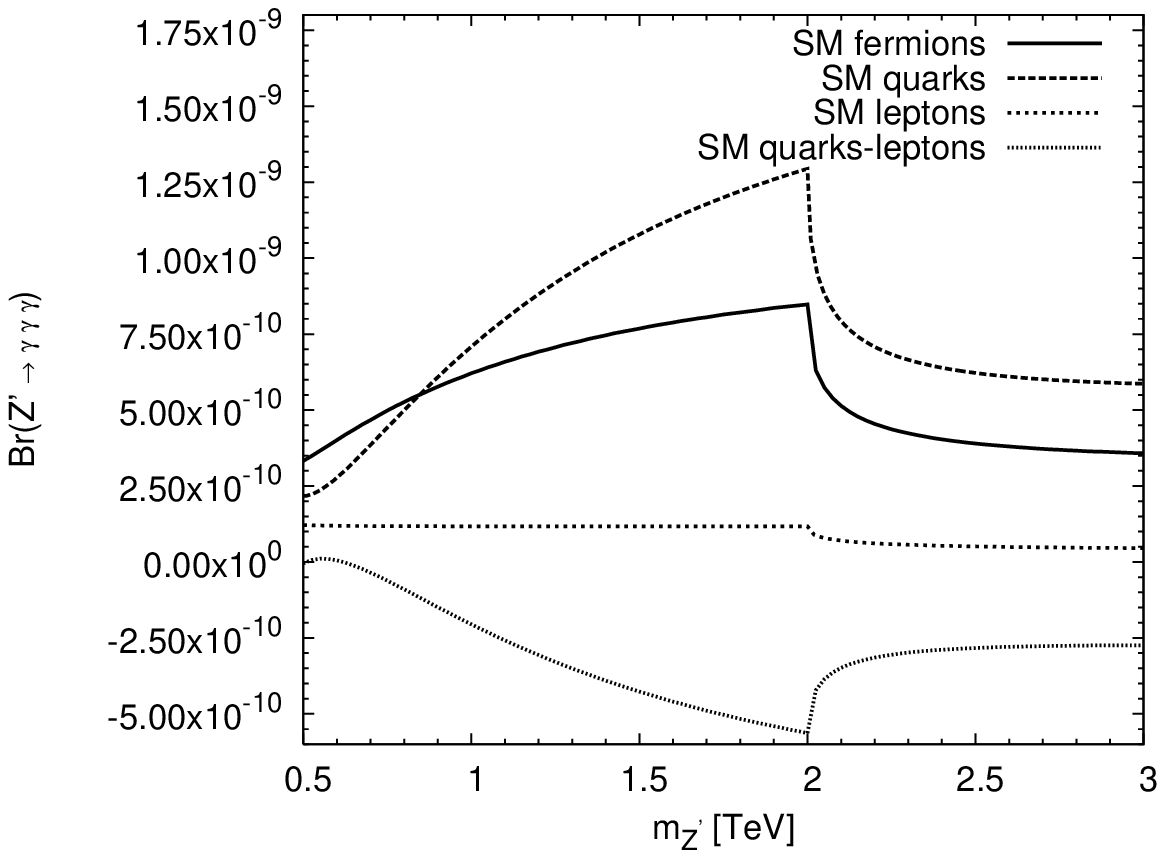}
\caption{\label{F5}Contribution of the SM fermions to  $Br(Z'\to \gamma \gamma \gamma)$ as a function of $m_{Z'}$. The left (right)-handed graph corresponds to a scenario with $m_Q=500$ GeV ($m_Q=1000$ GeV). The lepton, quark, interference, and total contributions are separately shown.}
\end{center}
\end{figure}

Having discussed the relative importance of the diverse spin contributions to the $Z'\to \gamma \gamma \gamma$ decay, we turn to discuss the corresponding branching ratio. To determine it, we need the main decay widths of the $Z'$ predicted by the minimal 331 model, which are given in Ref.~\cite{T2}. As already commented, a good strategy for studying the sensitivity of this decay to physics lying at the $m_{Z'}$ scale consists in assuming a degenerate spectrum for each type of new particles, \textit{i.e.}, we will assume that the three new exotic quarks are mass degenerate, the same will be assumed for the case of the four simply charged and three doubly charged Higgs bosons. The pairs of bileptons $Y^{\pm \pm}$ and $Y^{\pm}$ also will be taken with the same mass. This assumption is quite reasonable, as all these new particles receive their mass at the firs stage of spontaneous symmetry breaking. The contribution to the $Z'\to \gamma \gamma \gamma$ decay of the exotic quarks and the SM fermions are shown in Figs. \ref{F4} and \ref{F5}, respectively. From these figures, it can be appreciated that the exotic quarks  contribution ranges from $10^{-7}$ to $4\times 10^{-7}$ for $0.1<m_Q/m_{Z'}<0.5$, which is about three orders of magnitude larger than the combined contribution of the SM fermions. This is a surprising result, as it seems contradicts our previous analysis concerning the spin $1/2$ contribution to this decay, in which we concluded that the contributions of the lightest SM fermions dominate. This apparent contradiction is due to the fact that the global factor involving products of coupling constants (see Eq. (\ref{FermionicConstant})) is much larger in the case of exotic quarks than in the case of SM fermions, as it is shown in Table~\ref{TFC}.

\begin{table}[!ht]
\centering
\caption{\label{TFC} Relative importance of global factor given by Eq. (\ref{FermionicConstant}).}
\begin{tabular}{cr}\hline\hline
$F$    & $\big(g_{1/2}^F\big)^2$ \\ \hline
$l$    & 0.096 \\
$u, c$ & 0.221  \\
$d, s$ & 0.023 \\
$D, S$ & 441.715   \\
$b$    & 0.007 \\
$t$    & 5.475    \\
$T$    & 3435.560   \\\hline\hline
\end{tabular}
\end{table}

\begin{figure}[!ht]
\begin{center}
\includegraphics[width=8cm]{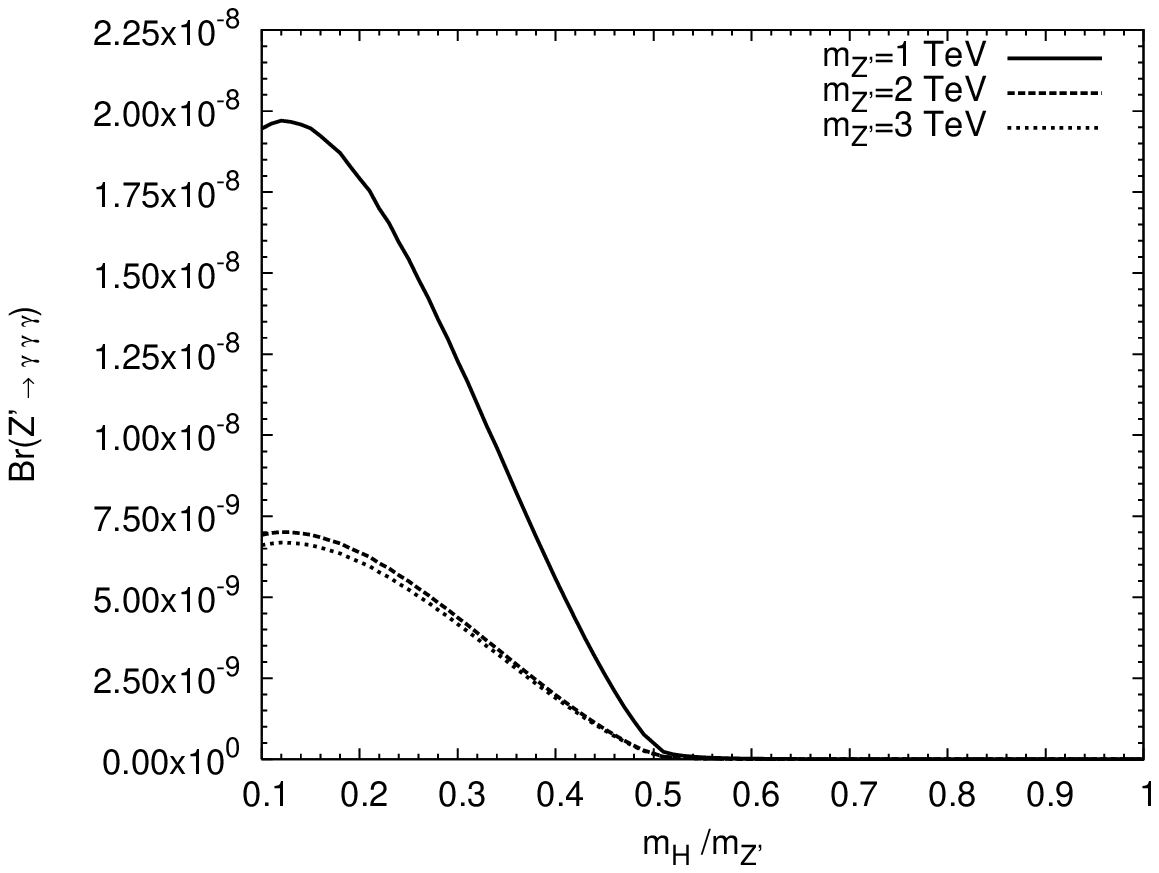}
\includegraphics[width=8cm]{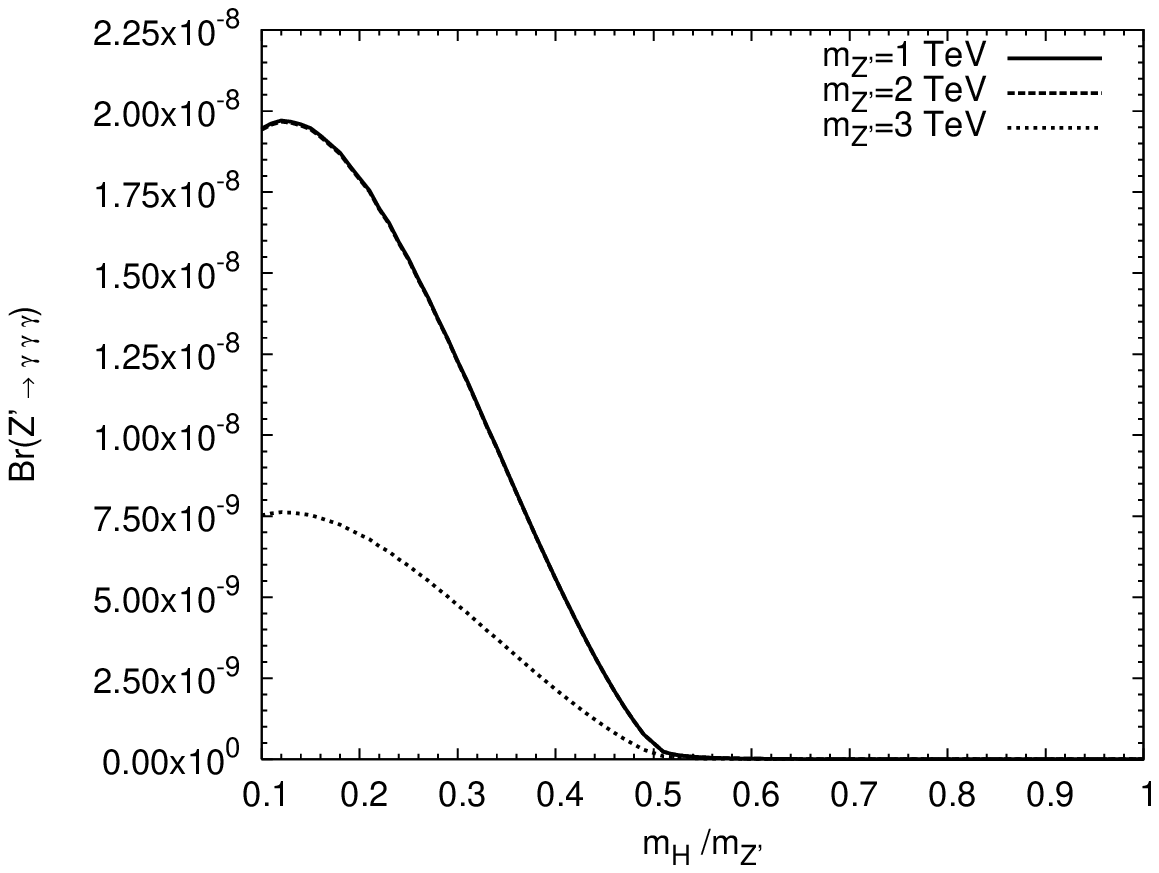}
\caption{\label{F6} Contribution of charged scalars to the $Br(Z'\to \gamma \gamma \gamma)$ as a function of $m_H/m_{Z'}$ for $m_Q=500$ GeV (left-handed) and $m_Q=1000$ GeV (right-handed).}
\end{center}
\end{figure}

We now turn to discuss the contribution of the charged scalars to the $Br(Z'\to \gamma \gamma \gamma)$. Its behavior as a function of $m_H/m_{Z'}$ is shown in Fig. \ref{F6}. From this figure, it can be appreciated that $Br(Z'\to \gamma \gamma \gamma)$ is larger for lowest values of $m_H/m_{Z'}$, which is consistent with our previous analysis for spin $0$ amplitudes shown in Fig. \ref{F2}. The contribution to the branching ratio can be of order of $10^{-8}$, at the best. This contribution to the $Z'\to \gamma \gamma \gamma$ decay is one order of magnitude lower than the exotic quarks one and, as it can be appreciated from Table~\ref{TSC}, the exotic doubly charged scalars dominate.

\begin{table}[!ht]
\centering
\caption{\label{TSC}  Relative importance of global factor given by Eq. (\ref{HiggsConstant}).}
\begin{tabular}{cr}\hline\hline
$H$           & $\big(g_0^H\big)^2$ \\ \hline
$h_1^+,h_4^+$ & 1.025 \\
$h_2^+$       & 0 \\
$h_3^+$       & 0.028 \\
$d_1^{++}$    & 130.435 \\
$d_2^{++}$    & 167.956 \\
$d_3^{++}$    & 65.608 \\ \hline\hline
\end{tabular}
\end{table}

\begin{figure}[!ht]
\begin{center}
\includegraphics[width=8cm]{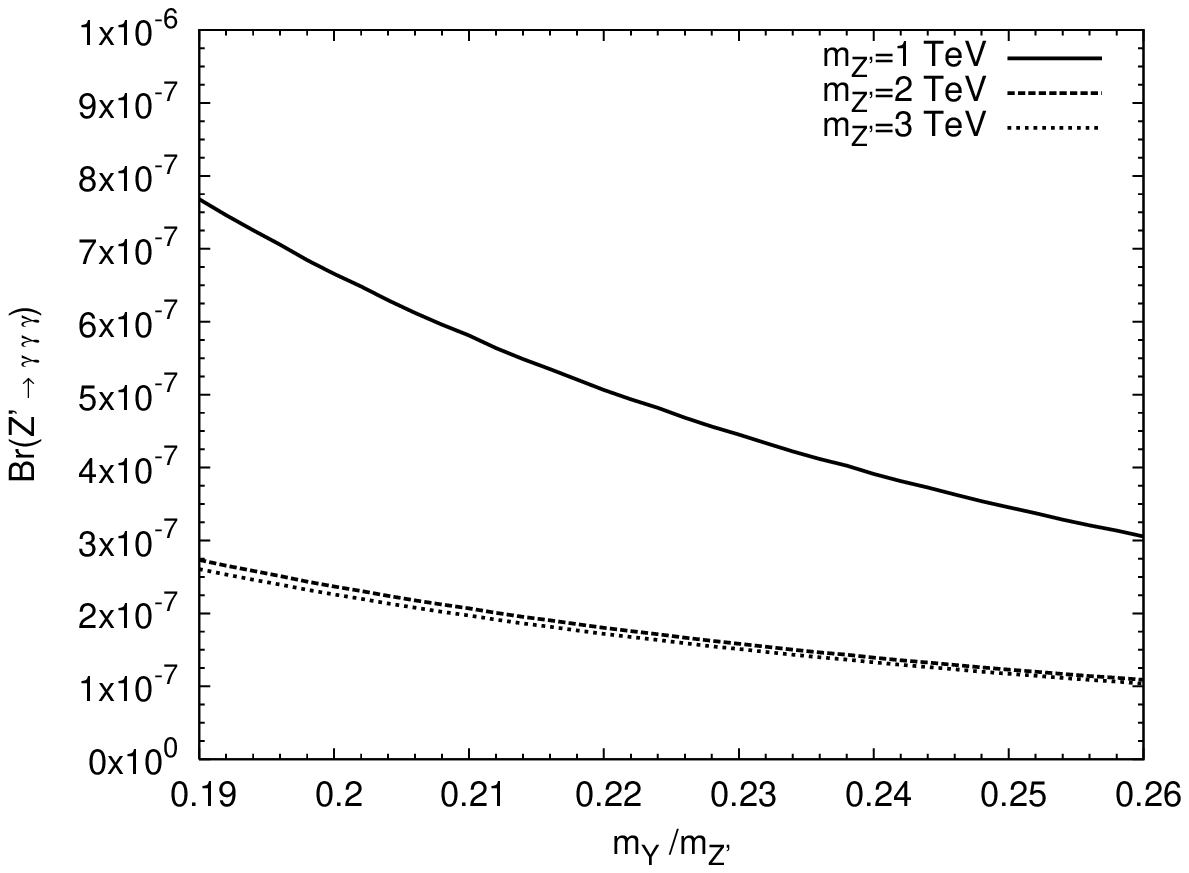}
\includegraphics[width=8cm]{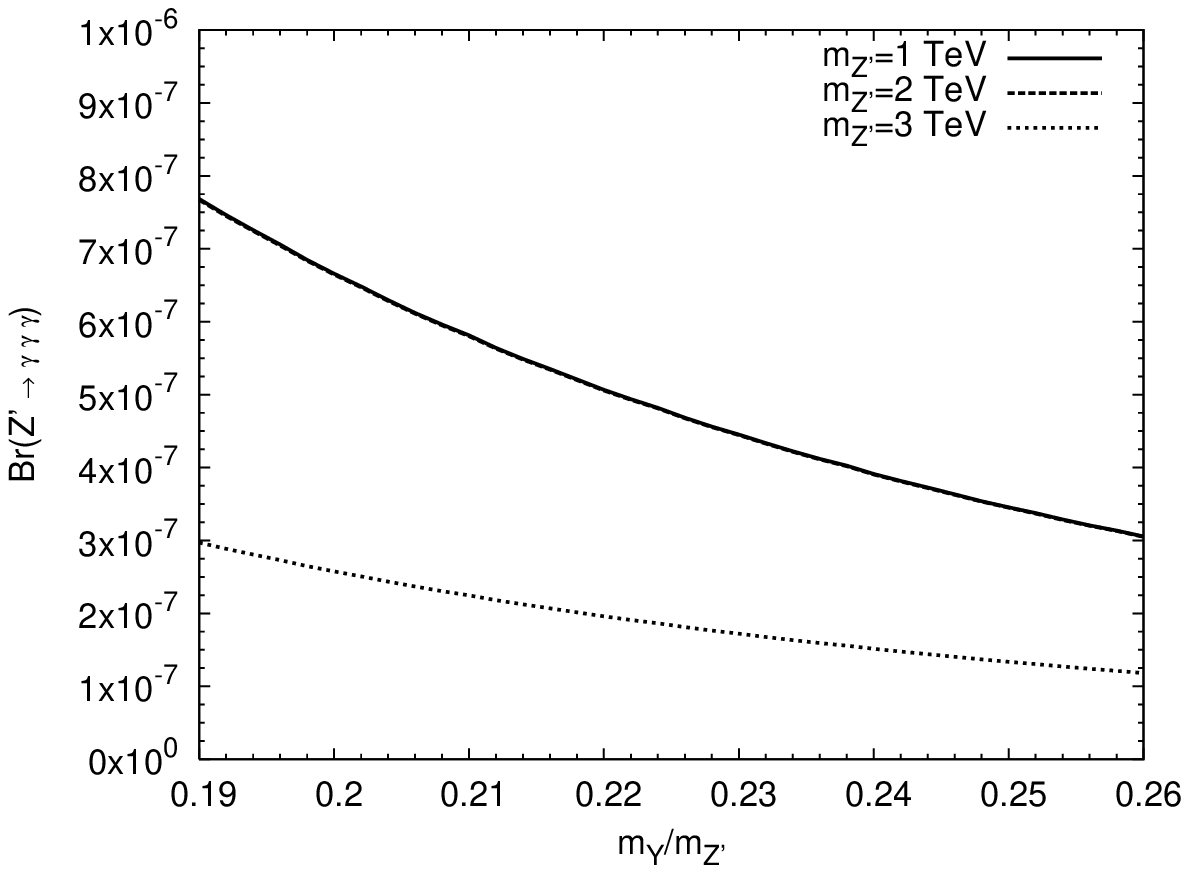}
\caption{\label{F7} Contribution of bilepton gauge bosons to $Br(Z'\to \gamma \gamma \gamma)$ as a function of $m_Y/m_Z'$ for $m_Q=500$ GeV (left-handed) and $m_Q=1000$ GeV (right-handed).}
\end{center}
\end{figure}

As far as the bilepton contribution to the $Z'\to \gamma \gamma \gamma$ decay is concerned, the behavior of the branching ratio as a function of $m_Y/m_{Z'}$ is shown in Fig. \ref{F7}. From this figure, it can be seen that the bilepton contribution to $Br(Z'\to \gamma \gamma \gamma)$ ranges, in the variation for $m_Y/m_{Z'}$ allowed by theoretical and experimental constraints, from approximately $8\times 10^{-7}$ to $3\times 10^{-7}$ (see Tables~\ref{TBR1} and \ref{TBR2}), at the best. This contribution is larger than the exotic quark one by approximately a factor of $2$. As it occurs in the case of exotic quarks, the exotic charge content of one of the bileptons play a decisive role in obtaining this result, as it can be appreciated from Table~\ref{TBC}.

\begin{table}[!ht]
\centering
\caption{\label{TBR1} Some values of $Br(Z'\to \gamma \gamma \gamma)$ in the scenario $m_Q=500$ GeV.}
\begin{tabular}{lccccc}\hline\hline
         &\vline&                 & $m_Y/m_{Z'}$ &              \\
Br       &\vline& 0.19            & & 0.26                      \\ \hline
$m_{Z'}$=1 TeV && 7.68$\times 10^{-7}$ & & 3.05$\times 10^{-7}$ \\
$m_{Z'}$=2 TeV && 2.73$\times 10^{-7}$ & & 1.09$\times 10^{-7}$ \\
$m_{Z'}$=3 TeV && 2.61$\times 10^{-7}$ & & 1.04$\times 10^{-7}$ \\ \hline\hline
\end{tabular}
\end{table}

\begin{table}[!ht]
\centering
\caption{\label{TBR2} Some values of $Br(Z'\to \gamma \gamma \gamma)$ in the scenario $m_Q=1000$ GeV.}
\begin{tabular}{lccccc}\hline\hline
         &\vline&                      & $m_Y/m_{Z'}$ &                       \\
Br       &\vline& 0.19                 &              & 0.26                  \\ \hline
$m_{Z'}$=1 TeV && 7.68$\times 10^{-7}$ &              & 3.06$\times 10^{-7}$  \\
$m_{Z'}$=2 TeV && 7.67$\times 10^{-7}$ &              & 3.05$\times 10^{-7}$  \\
$m_{Z'}$=3 TeV && 2.97$\times 10^{-7}$ &              & 1.18$\times 10^{-7}$  \\ \hline\hline
\end{tabular}
\end{table}

\begin{table}[!ht]
\centering
\caption{\label{TBC} Relative importance of global factor given by Eq. (\ref{BosonicConstant}).}
\begin{tabular}{cr}\hline\hline
$V$      & $\big(g_1^V\big)^2$ \\ \hline
$Y^+$    & 0.226 \\
$Y^{++}$ & 14.469 \\\hline\hline
\end{tabular}
\end{table}

\begin{figure}[!ht]
\begin{center}
\includegraphics[width=8cm]{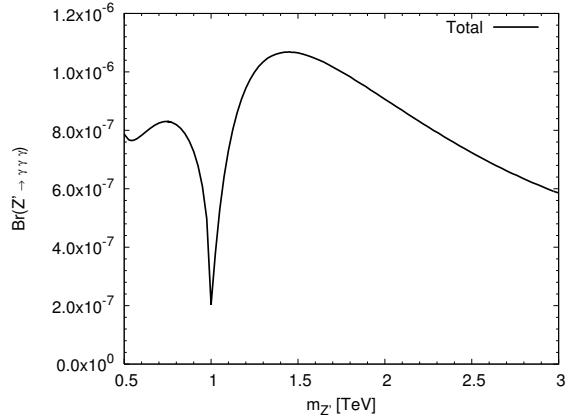}
\caption{\label{F8} Total contribution to $Br(Z'\to \gamma \gamma \gamma)$ for $m_Q=500$ GeV, $m_Y/m_{Z'}=0.19$ and $m_H=250$ GeV.}
\end{center}
\end{figure}

Finally, the total contribution to $Br(Z'\to \gamma \gamma \gamma)$ is displayed in Fig.~\ref{F8}. From this figure, it can be appreciated that the branching ratio ranges from $7.87\times 10^{-7}$ to $5.78\times 10^{-7}$ for 0.5 TeV$<m_Z'<$3 TeV, in a scenario with $m_Q=500$ GeV, $m_Y/m_Z'=0.19$, and $m_H=250$ GeV. As it is shown in Table~\ref{TTBR}, inside of this range of variation of $m_{Z'}$ the branching ratio can reach a maximum of $1.07\times 10^{-6}$, and a minimum of $2.03\times 10^{-7}$ value, which occur for $m_Z'=1.45$ TeV and 1 TeV, respectively. However, it can be appreciated from Fig. \ref{F8} that with the exception of a small interval centered in $m_{Z'}=1$ TeV, the branching ratio for the $Z'\to \gamma \gamma \gamma$ decay is essentially of $10^{-6}$. It is worth comparing this result with the branching ratios obtained in reference~\cite{T2} for the rare one-loop decays $Z'\to ZZ$ and $Z'\to Z\gamma$, which are of the order of $10^{-6}$ and $10^{-10}$, respectively. This shows that the $Z'\to ZZ$ and $Z'\to \gamma \gamma \gamma$ decays have branching ratios of the same order of magnitude in the minimal 331 model. This surprising result can be explained by noting that the $Z'\to ZZ$ decay only receives contributions from the fermionic sector, whereas the $Z'\to \gamma \gamma \gamma$ one receives contributions from both the fermionic and bosonic sectors of the model. In addition, as already seen, the exotic charge content of the new quarks and gauge bosons play a central role in the three body decay.

\begin{table}[!ht]
\centering
\caption{\label{TTBR} Some values of the total $Br(Z'\to \gamma \gamma \gamma)$ for $m_Q=500$ GeV,  $m_Y/m_{Z'}=0.19$, and $m_H=250$ GeV (see Fig. \ref{F8}). The maximum value, $1.07\times 10^{-6}$, occurs at $m_{Z'}=1.45$ TeV.}
\begin{tabular}{lcccc}\hline\hline
      &\vline&                      & $m_{Z'}$ [TeV]       &                      \\
Br    &\vline& 0.5                  & 1                    & 3                    \\ \hline
Total &      & 7.87$\times 10^{-7}$ & 2.03$\times 10^{-7}$ & 5.78$\times 10^{-7}$ \\ \hline\hline
\end{tabular}
\end{table}

\section{Conclusions}
\label{c}Purely photonic decays of self-conjugate vector fields, to which the particle coincides with its antiparticle, as it is the case of the $V^0=Z,Z'$
gauge bosons considered in this work, are very constrained by gauge invariance, Bose statistics, and Lorentz invariance. Gauge invariance forbids any coupling of $V^0$ to photons at the tree level, so they only can arise as a quantum fluctuation of one-loop or higher orders. Since couplings of $V^0$ to two photons cannot exist due to the Landau-Yang theorem~\cite{LY}, the interaction with three photons is the most important electromagnetic coupling of $V^0$, which, in the context of a renormalizable theory, first arise at the one-loop level. Gauge invariance restricts this coupling to be characterized by dimension-six operators of the form $(f/m^4)V^0_{\mu \nu}F^{\lambda \nu}F_{\lambda \rho}F^{\mu \rho}$~\cite{PTT}, where $f$ represents a loop amplitude and $m$ is the mass of the particle circulating in the loop. From the decoupling theorem~\cite{DT}, one expects that if $m\gg m_{V^0}$, the loop effect of the heavy particle decouples quickly. In contrast, one expect a relevant contribution if $m<m_{V^0}$ and it is more and more important if $m$ is smaller than $m_{V^0}$. In this work, we have studied the decays of $Z$ and $Z'$ into three photons within the context of the minimal 331 model, which predicts the existence of three new exotic quarks, two new gauge bosons, one simply charged and one doubly charged, and four simply charged and three doubly charged scalars. All the features of the $V^0\gamma \gamma \gamma$ coupling commented above were reproduced. In the case of the SM $Z$ gauge boson, it was found that the $Z\to \gamma \gamma \gamma$ decay is insensitive to new physics effects, as the masses of the new particles are much larger than $m_Z$. Although large global factors arising from exotic charge contents of the new particles can substantially increase the loop amplitude, the well-known branching fraction of about $10^{-9}$, which is determined essentially by the lightest fermions and remains unchanged. Due to its insensitivity to heavy physics effects, this decay will likely be beyond the reach of the LHC or the future ILC. As far as the $Z'\to \gamma \gamma \gamma$ decay is concerned, it was found that it can have a branching fraction as large as $10^{-6}$, which may be at the reach of future colliders. In particular, in the case of the LHC it has been found~\cite{Langacker1} that the primary discovery mode for a $Z'$ boson is a dilepton resonance via the Drell-Yan production process with a branching fraction of order $10^{-2}$ for a $Z'$ mass in the range of the TeV scale and with an integrated luminosity of 100 to 300 $fb^{-1}$. Our results for the branching fraction for $Z'\to \gamma \gamma \gamma$ thus makes it rather difficult to detect this decay mode at the LHC. In accordance with the previous general discussion, this decay is more important in the measure that the masses of the particles circulating in the loop turn out to be smaller than $m_{Z'}$. It was found that the exotic quarks and bilepton gauge bosons contribution is the dominant one, whereas the scalar contribution is smaller by about one order of magnitude. It was found that the exotic charge content of the particles circulating in the loop play a crucial role in this decay. In general terms, our study shows us that the decay of a self-conjugate vector boson $V^0$ into three photons is favored by three circumstances: 1) the presence of particles with exotic content of charge, 2) such particles have masses substantially smaller than $m_{V^0}$, and 3) there are several exotic particles.

\acknowledgments{We acknowledge financial support from CONACYT and SNI (M\' exico).}

\appendix

\section{Feynman rules}\label{FR}
In this Appendix, we present the Feynman rules used in the paper.

\begin{figure}[!ht]
\begin{center}
\includegraphics[width=7cm]{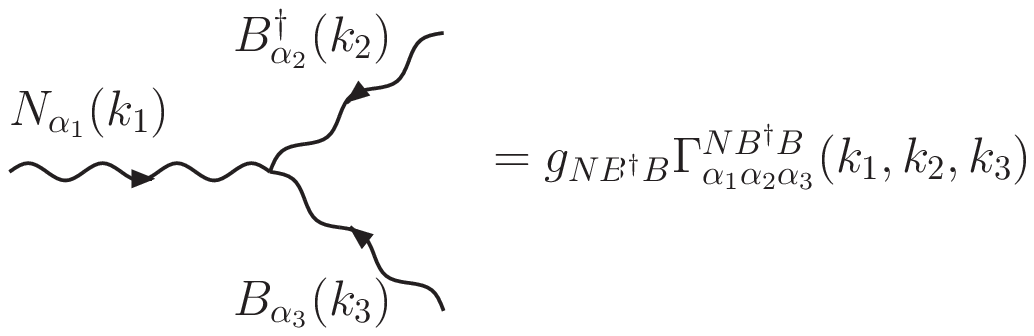} \ \
\includegraphics[width=6cm]{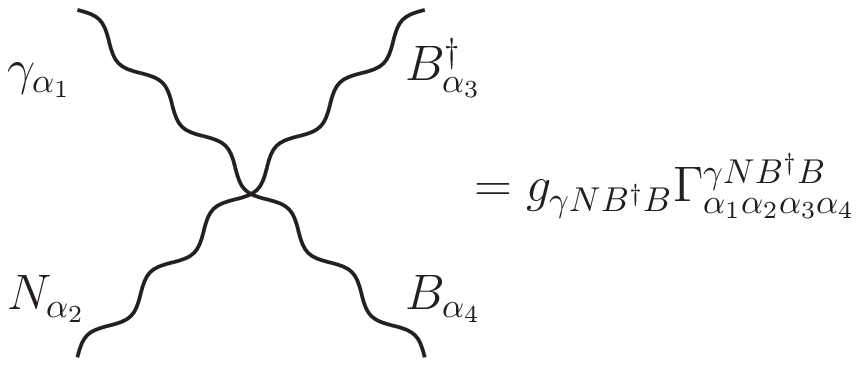}
\includegraphics[width=7cm]{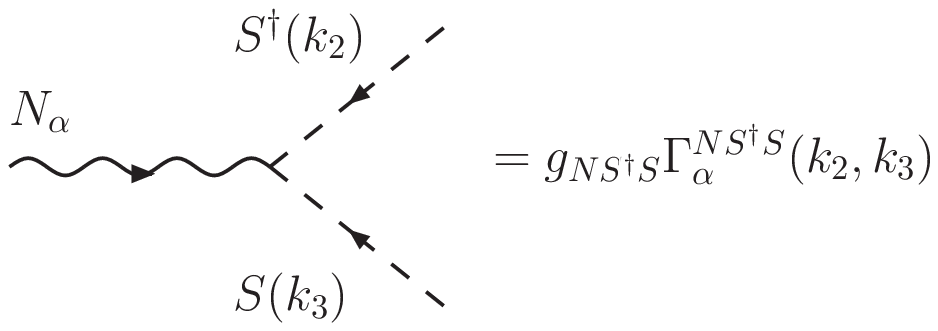} \ \
\includegraphics[width=6cm]{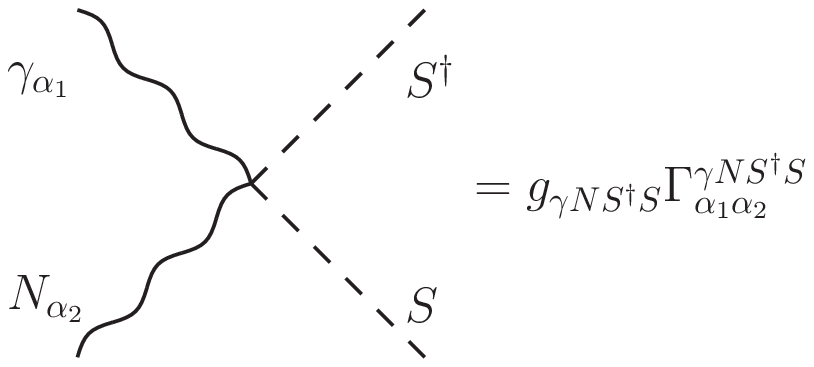}
\caption{Bosonic Feynman rules needed for the calculation of the $V^0\to \gamma \gamma \gamma$ decay. In this figures, $N\equiv$ $\gamma,\; V^0$, with $V^0\equiv$ $Z,\; Z'$; $B$=$W^+$, $Y^+$, $Y^{++}$; $S$=$G_W^+$, $C_W^+$, $\bar{C}_W^+$, $G_Y^+$, $C_Y^+$, $\bar{C}_Y^+$, $G_Y^{++}$, $C_Y^{++}$, $\bar{C}_Y^{++}$; $S$=$H$=$h_1^+$, $h_2^+$, $h_3^+$, $h_4^+$, $d_1^{++}$, $d_2^{++}$, $d_3^{++}$.}\label{rules}
\end{center}
\end{figure}

In Fig.~\ref{rules} $g_{\gamma B^\dag B}=-ieQ_B$, $g_{V^0B^\dag B}=-igg_{V^0}^B/2c_W$, $g_{\gamma\gamma B^\dag B}=-ie^2Q_B^2$,
$g_{\gamma V^0B^\dag B}=-ieQ_Bgg_{V^0}^B/2c_W$, $g_{\gamma S^\dag S}=ieQ_S$, $g_{V^0S^\dag S}=igg_{V^0}^S/2c_W$, $g_{\gamma\gamma S^\dag S}=ie^2Q_S^2$, and  $g_{\gamma V^0S^\dag S}=ieQ_Sgg_{V^0}^S/2c_W$. On the other hand, the tensorial functions are given by
\begin{eqnarray}
\Gamma_{\alpha_1\alpha_2\alpha_3}^{NB^\dag B}(k_1,k_2,k_3)&=&
    (k_2-k_3)_{\alpha_1} g_{\alpha_2\alpha_3}
    +\bigg(-k_1-\frac{\alpha_{NB^\dag B}}{\xi}k_2+k_3\bigg)_{\alpha_2}g_{\alpha_1\alpha_3} \nonumber\\
&& +\bigg(k_1-k_2+\frac{\alpha_{NB^\dag B}}{\xi}k_3\bigg)_{\alpha_3}g_{\alpha_1\alpha_2}, \\
\Gamma_{\alpha_1\alpha_2\alpha_3\alpha_4}^{\gamma NB^\dag B}&=&
    2g_{\alpha_1\alpha_2}g_{\alpha_3\alpha_4}
    -\bigg(1+\frac{\alpha_{\gamma NB^\dag B}}{\xi}\bigg)(g_{\alpha_1\alpha_3}g_{\alpha_2\alpha_4}+
    g_{\alpha_1\alpha_4}g_{\alpha_2\alpha_3}), \\
\Gamma_\alpha^{NS^\dag S}(k_2,k_3)&=&(k_2-k_3)_\alpha, \\
\Gamma_{\alpha_1\alpha_2}^{\gamma NS^\dag S}&=& 2 g_{\alpha_1\alpha_2},
\end{eqnarray}
 where $\alpha_{ZW^-W^+}=\alpha_{\gamma ZW^-W^+}=t_W^2$ and $\alpha_{\gamma B^\dag B}=\alpha_{\gamma\gamma B^\dag B}=\alpha_{V^0Y^\dag Y}=\alpha_{\gamma V^0Y^\dag Y}=-1$.

The diverse factors appearing in Sec. \ref{a} concerning the couplings of the $Z'$ gauge boson to pairs of fermions, charges scalars, and  bileptons are given in tables \ref{TCF}, \ref{TCS}, and \ref{TCB}, respectively.

\begin{table}[!ht]
\centering
\caption{\label{TCF} Couplings of $Z$ and $Z'$ gauge bosons to pairs of SM and exotic fermions. Only the vectorial component is shown.}
\begin{tabular}{cccc}\hline\hline
$f$          & $Q_f$         & $g_{VZ}^f$            & $g_{VZ'}^f$ \\ \hline
$e,\mu,\tau$ & $-1$             & $-\frac{1-4s_W^2}{2}$ & $\frac{\sqrt{3}\sqrt{1-4s_W^2}}{2c_W^2}$ \\
$u,c$        & $\frac{2}{3}$ & $\frac{3-8 s_W^2}{6}$ & $-\frac{1-6s_W^2}{2\sqrt{3}\;c_W^2\sqrt{1-4s_W^2}}$ \\
$d,s$        & $-\frac{1}{3}$ & $-\frac{3-4s_W^2}{6}$ & $-\frac{1}{2\sqrt{3}\;c_W^2\sqrt{1-4s_W^2}}$ \\
$D,S$        & $-\frac{4}{3}$ & $\frac{8 s_W^2}{3}$   & $\frac{1-9s_W^2}{\sqrt{3}\;c_W^2\sqrt{1-4s_W^2}}$ \\
$b$          & $-\frac{1}{3}$ & $-\frac{3-4s_W^2}{6}$ & $\frac{1-2s_W^2}{2\sqrt{3}\;c_W^2\sqrt{1-4s_W^2}}$ \\
$t$          & $\frac{2}{3}$ & $\frac{3-8s_W^2}{6}$  & $\frac{1+4s_W^2}{2\sqrt{3}\;c_W^2\sqrt{1-4s_W^2}}$ \\
$T$          & $\frac{5}{3}$ & $-\frac{10 s_W^2}{3}$ & $-\frac{1-11s_W^2}{\sqrt{3}\;c_W^2\sqrt{1-4s_W^2}}$ \\ \hline\hline
\end{tabular}
\end{table}
\begin{table}[!ht]
\centering
\caption{\label{TCS} Couplings of $Z$ and $Z'$ gauge bosons to pairs of charged scalars. The value $\theta\sim$ 0 is assumed for the $Z'-Z$ mixing angle. Also the values $|\lambda_i|$, $|\tilde{f}_j|$ $\sim$ 1 are assumed.}
\begin{tabular}{cccc}\hline\hline
$S$           & $Q_S$ & $g_Z^S$     & $g_{Z'}^S$                \\ \hline
$G_W^+$       & 1     & $c_{2W}$    & $-\frac{c_Wc_\theta}{\sqrt{3}}t_{\theta}^2$   \\
$G_Y^+$       & 1     & $-1-2s_W^2$ & $\frac{c_Wc_\theta}{\sqrt{3}}(1-2t_\theta^2)$ \\
$G_Y^{++}$    & 2     & $1-4s_W^2$  & $\frac{c_Wc_\theta}{\sqrt{3}}(1-2t_\theta^2)$ \\
$h_1^+,h_4^+$ & 1     & $-2s_W^2$   & $\frac{2c_Wc_\theta}{\sqrt{3}}$ \\
$h_2^+$       & 1     & $c_{2W}$    & $\frac{c_Wc_\theta}{\sqrt{3}}[1-2(1+t_\theta^2)H_{22}^2]$ \\
$h_3^+$       & 1     & $c_{2W}$    & $\frac{c_Wc_\theta}{\sqrt{3}}[1-2(1+t_\theta^2)H_{32}^2]$ \\
$d_1^{++}$    & 2     & $-4s_W^2$   & $\frac{2c_Wc_\theta}{\sqrt{3}}\widetilde{N}_4^2[2\widetilde{X}_4^2+(1-t_\theta^2)\tilde{a}^2]$ \\
$d_2^{++}$    & 2     & $-4s_W^2$   & $\frac{2c_Wc_\theta}{\sqrt{3}}\widetilde{N}_5^2[2\widetilde{X}_5^2+(1-t_\theta^2)\tilde{a}^2]$ \\
$d_3^{++}$    & 2     & $2c_{2W}$   & $\frac{2c_Wc_\theta}{\sqrt{3}}$ \\ \hline\hline
\end{tabular}
\end{table}
\begin{table}[!ht]
\centering
\caption{\label{TCB} Couplings of $Z$ and $Z'$ to pairs of charged gauge bosons.}
\begin{tabular}{cccc}\hline\hline
$B$      & $Q_B$ & $g_Z^B$     & $g_{Z'}^B$                \\ \hline
$W^+$    & 1     & $2c_W^2$    & 0                         \\
$Y^+$    & 1     & $-1-2s_W^2$ & $\sqrt{3}\sqrt{1-4s_W^2}$ \\
$Y^{++}$ & 2     & $1-4s_W^2$  & $\sqrt{3}\sqrt{1-4s_W^2}$ \\ \hline\hline
\end{tabular}
\end{table}

\newpage

\section{Form factors}\label{FFactors}
The form factors associated with the spin $1/2$ contributions are given in Ref.~\cite{T4}. Here we present the form factors induced by the bosonic particles of the model. The spin $1$ form factors are given by
\begin{eqnarray}
    f_{B1} &=& \frac{B_0^B(1) (p_{13}-2 p_{23}) \alpha}{12 p_{13}^2 p_{23}}
    +\frac{B_0^B(2) (p_{12}+p_{23}) \alpha}{12 p_{12}^2 p_{23}}
    -\frac{B_0^B(3) (2 p_{12}-p_{13}) p_{23} \alpha}{12 p_{12}^2 p_{13}^2} \nonumber\\
    && -\frac{B_0^B(4) (p_{12}+p_{13}+p_{23}) [p_{12} (p_{13}-2 p_{23})+p_{13} p_{23}] \alpha}{12
    p_{12}^2 p_{13}^2 p_{23}} \nonumber\\
    && +\frac{C_0^B(1) \{[\alpha(p_{13}^3-2 p_{23}^3)+12 p_{23}^2 p_{13}]
    p_{12}^2+3 m_B^2 p_{13}^2 p_{23}^2 \alpha\} }{12 p_{12} p_{13}^3 p_{23}^2}
    +\frac{C_0^B(2) p_{13} [\alpha(p_{12}^3+p_{23}^3)-12 p_{23}^2 p_{12}]}{12
    p_{12}^3 p_{23}^2} \nonumber\\
    && +\frac{C_0^B(3) p_{23} [2 (6 p_{13}-p_{23} \alpha) p_{12}^3+3 p_{13}^2 (m_B^2 \alpha-4 p_{13}) p_{12}+p_{13}^3 p_{23} \alpha]}{12 p_{12}^3 p_{13}^3} \nonumber\\
    && -\frac{C_0^B(4) (p_{13}+p_{23}) (\alpha p_{13}^3+12 p_{23}^2 p_{13}-2 p_{23}^3 \alpha )}{12 p_{13}^3 p_{23}^2} \nonumber\\
    && -\frac{C_0^B(5) (p_{12}+p_{23}) [\alpha(p_{13}p_{12}^3+p_{13} p_{23}^3)+3 p_{23}^2 (m_B^2 \alpha-4 p_{13}) p_{12}]}{12 p_{12}^3 p_{13} p_{23}^2} \nonumber\\
    && -\frac{C_0^B(6) (p_{12}+p_{13}) [2 (6 p_{13}-p_{23} \alpha) p_{12}^3-12 p_{13}^3 p_{12}+p_{13}^3
    p_{23} \alpha]}{12 p_{12}^3 p_{13}^3} \nonumber\\
    && -\frac{D_0^B(1) [2 p_{12} p_{23} (12 p_{23}+p_{13} \alpha) m_B^2+p_{23}^2 (2 m_B^2\alpha-3 p_{13}\beta) m_B^2+2 p_{12}^2 p_{13}^2 \alpha] }{12 p_{12}
    p_{13} p_{23}^2} \nonumber\\
    && -\frac{D_0^B(2) \{2 p_{13}^2 \alpha m_B^4+p_{13} [4 p_{12} (6 p_{13}-p_{23} )+3 p_{13} (p_{23}
    \alpha-8 p_{13})] m_B^2+4 p_{12}^2 p_{23} (6 p_{13}-p_{23} \alpha)\}}{12 p_{12} p_{13}^3} \nonumber\\
    && -\frac{D_0^B(3) [2 p_{12}^2 \alpha m_B^4+p_{12} (24 p_{12}^2-24 p_{13} p_{12}+5 p_{13}
    p_{23} \alpha) m_B^2+2 p_{13}^2 p_{23} (p_{23} \alpha-12 p_{12})]}{12 p_{12}^3 p_{13}} \nonumber\\
    && +\frac{\alpha}{12 p_{12} p_{13}} \ ,
\end{eqnarray}

\begin{eqnarray}
    f_{B7} &=& \frac{B_0^B(1) \alpha}{2 p_{23}^3}
    +\frac{B_0^B(2) [p_{12} (2 p_{13}+p_{23})+p_{23} (3 p_{13}+p_{23})] \alpha}{4 p_{23}^3 (p_{12}+p_{23})^2} -\frac{B_0^B(4) (p_{12}+p_{13}+p_{23}) (2 p_{12}+3 p_{23}) \alpha}{4 p_{23}^3 (p_{12}+p_{23})^2} \nonumber\\
    && +\frac{C_0^B(1) (2 p_{13}+p_{23}) (p_{23} m_B^2+p_{12} p_{13}) \alpha}{4 p_{13}p_{23}^4} \nonumber\\
    && +\frac{C_0^B(2) [p_{13} p_{23}^2 m_B^2+p_{12} p_{23} (2 p_{13}+p_{23}) m_B^2+p_{12}^2 p_{13}
    (2 p_{13}+p_{23})] \alpha}{4 p_{12}^2 p_{23}^4}
    +\frac{C_0^B(3) m_B^2 \alpha}{4 p_{12}^2 p_{23}} \nonumber\\
    && -\frac{C_0^B(4) (p_{13}+p_{23}) [2 p_{23} m_B^2+p_{12} (2 p_{13}+p_{23})] \alpha}{4p_{12} p_{23}^4} \nonumber\\
    && -\frac{C_0^B(5)[(2 p_{13}+p_{23}) p_{12}^4+2 p_{23} (m_B^2+2 p_{13}+p_{23}) p_{12}^3+p_{23}^2
    (3 m_B^2+2 p_{13}+p_{23}) p_{12}^2+m_B^2 p_{23}^4]\alpha}{4 p_{12}^2 p_{23}^4
    (p_{12}+p_{23})} \nonumber\\
    && -\frac{C_0^B(6) m_B^2 (p_{12}+p_{13})^2 \alpha}{4 p_{12}^2 p_{13} p_{23}^2}
    -\frac{D_0^B(1) [2 p_{23}^2 m_B^4+p_{12} p_{23} (8 p_{13}+3 p_{23}) m_B^2+2 p_{12}^2 p_{13}
    (2 p_{13}+p_{23})] \alpha}{4 p_{12} p_{23}^4} \nonumber\\
    && -\frac{D_0^B(2) m_B^2 (2 p_{13} m_B^2+p_{12} p_{23}) \alpha}{4 p_{12} p_{13}p_{23}^2}
    -\frac{D_0^B(3) m_B^2 [2 p_{12} m_B^2+(p_{12}+2 p_{13}) p_{23}] \alpha}{4 p_{12}^2 p_{23}^2}
    -\frac{\alpha}{4 p_{23}^2 (p_{12}+p_{23})} \ ,
\end{eqnarray}

\begin{eqnarray}
    f_{B13} &=& \frac{B_0^B(1) [8 p_{12} p_{13}^3+3 (4 p_{12}+p_{13}) p_{23} p_{13}^2-6 p_{12} p_{23}^2 p_{13}-(4
    p_{12}+3 p_{13}) p_{23}^3] \alpha}{24 p_{13}^2 p_{23}^2 (p_{13}+p_{23})^2} \nonumber\\
    && +\frac{B_0^B(2) (8 p_{12}-p_{23}) \alpha}{24 p_{12} p_{23}^2}
    -\frac{B_0^B(3) (4 p_{12}+p_{13}) \alpha}{24 p_{12} p_{13}^2} \nonumber\\
    && -\frac{B_0^B(4) (p_{12}+p_{13}+p_{23}) [2 p_{12} (4 p_{13}^3+6 p_{23} p_{13}^2-3 p_{23}^2
    p_{13}-2 p_{23}^3)-p_{13} p_{23} (p_{13}+p_{23})^2] \alpha}{24 p_{12} p_{13}^2 p_{23}^2
    (p_{13}+p_{23})^2} \nonumber\\
    && -\frac{C_0^B(1) p_{12} \{4 p_{12} (2 p_{13}^3-p_{23}^3) \alpha+3 p_{13} p_{23}
    [\alpha(2 p_{13} m_B^2+p_{13}^2)-p_{23}^2 \beta]\}}{24 p_{13}^3
    p_{23}^3} \nonumber\\
    && +\frac{C_0^B(2) [8 p_{13} p_{12}^3+3 (2 m_B^2+p_{13}) p_{23} p_{12}^2-p_{13} p_{23}^3]
    \alpha}{24 p_{12}^2 p_{23}^3}
    -\frac{C_0^B(3) p_{23} [\alpha(4 p_{12}^3+p_{13}^3)+3 p_{13} \beta p_{12}^2]}{24 p_{12}^2 p_{13}^3} \nonumber\\
    && +\frac{C_0^B(4)}{24 p_{13}^3 p_{23}^3 (p_{13}+p_{23})} \{ 3 p_{13} p_{23} [-2 p_{13} (p_{13}^2+2 p_{23} p_{13}-p_{23}^2) \alpha m_B^2-(p_{13}+p_{23})^2 (p_{13}^2 \alpha-p_{23}^2 \beta)]\nonumber\\
    &&-4 p_{12}(p_{13}+p_{23})^2 (2 p_{13}^3-p_{23}^3) \alpha \} \nonumber\\
    && -\frac{C_0^B(5) (p_{12}+p_{23}) [ p_{13} p_{12}^3+3 (2 m_B^2+p_{13}) p_{23} p_{12}^2-p_{13}
    p_{23}^3] \alpha}{24 p_{12}^2 p_{13} p_{23}^3} \nonumber\\
    && +\frac{C_0^B(6) (p_{12}+p_{13}) [\alpha(4p_{12}^3+p_{13}^3)
    +3 p_{13} \beta p_{12}^2]}{24 p_{12}^2 p_{13}^3} \nonumber\\
    && -\frac{D_0^B(1) \{\alpha[8 p_{12}^2p_{13}^2+p_{12} (14 m_B^2+3 p_{13}) p_{23}p_{13}]+p_{23}^2 [2 \alpha m_B^4+3 (p_{13} \alpha-8 p_{23}) m_B^2+12
    p_{13} p_{23}]\}}{12 p_{13} p_{23}^3} \nonumber\\
    && +\frac{D_0^B(2)}{12 p_{13}^3 p_{23}} [-2 p_{13}^2 \alpha m_B^4+p_{13} p_{23} (4 p_{12} \alpha+3 p_{13}
    \beta) m_B^2+p_{23} (-12 p_{13}^3+3 p_{12} p_{23} \beta p_{13}+4 p_{12}^2 p_{23}\alpha)]\nonumber\\
    && +\frac{D_0^B(3) [-2 p_{12}^2 \alpha m_B^4+p_{12} p_{23} (24 p_{12}+p_{13} \alpha ) m_B^2+p_{13} p_{23} (p_{13} p_{23} \alpha-12 p_{12}^2)]}{12 p_{12}^2 p_{13}p_{23}} \nonumber\\
    &&-\frac{(2 p_{13}-p_{23}) \alpha}{12 p_{13} p_{23} (p_{13}+p_{23})} \ .
\end{eqnarray}
In the above expressions, $B$ stands for $W^+$, $Y^+$ or $Y^{++}$. In addition, $\alpha\equiv \alpha_B-3$, $\beta\equiv \alpha_B+5$, $\alpha_W$=$t_W^2$, and $\alpha_Y$=$-1$.

On the other hand, the form factors associated with spin $0$ particles are given by
\begin{eqnarray}
    f_{S1} &=& -\frac{B_0^S(1) (p_{13}-2 p_{23})}{12 p_{13}^2 p_{23}}
    -\frac{B_0^S(2) (p_{12}+p_{23})}{12 p_{12}^2 p_{23}}
    +\frac{B_0^S(3) (2 p_{12}-p_{13}) p_{23}}{12 p_{12}^2 p_{13}^2} \nonumber\\
    && +\frac{B_0^S(4) (p_{12}+p_{13}+p_{23}) [p_{12} (p_{13}-2 p_{23})+p_{13} p_{23}]}{12 p_{12}^2 p_{13}^2p_{23}}
    -\frac{C_0^S(1) (p_{12}^2 p_{13}^3+3 m_S^2 p_{23}^2 p_{13}^2-2 p_{12}^2 p_{23}^3)}{12 p_{12}
    p_{13}^3 p_{23}^2} \nonumber\\
    && -\frac{C_0^S(2) p_{13} (p_{12}^3+p_{23}^3)}{12 p_{12}^3 p_{23}^2}
    +\frac{C_0^S(3) p_{23} (2 p_{23} p_{12}^3-3 m_S^2 p_{13}^2 p_{12}-p_{13}^3p_{23})}{12p_{12}^3p_{13}^3}\nonumber\\
    && +\frac{C_0^S(4) (p_{13}+p_{23}) (p_{13}^3-2 p_{23}^3)}{12 p_{13}^3 p_{23}^2}
    +\frac{C_0^S(5) (p_{12}+p_{23}) (p_{13} p_{12}^3+3 m_S^2 p_{23}^2 p_{12}+p_{13} p_{23}^3)}{12
    p_{12}^3 p_{13} p_{23}^2} \nonumber\\
    && -\frac{C_0^S(6) (p_{12}+p_{13}) (2 p_{12}^3-p_{13}^3) p_{23}}{12 p_{12}^3 p_{13}^3}
    +\frac{D_0^S(1) [2 p_{23}^2 m_S^4+p_{13} (2 p_{12}-3 p_{23}) p_{23} m_S^2+2 p_{12}^2 p_{13}^2]}{12
    p_{12} p_{13} p_{23}^2} \nonumber\\
    && +\frac{D_0^S(2) [2 p_{13}^2 m_S^4+p_{13} (3 p_{13}-4 p_{12}) p_{23} m_S^2-4 p_{12}^2 p_{23}^2]}{12
    p_{12} p_{13}^3} \nonumber\\
    && +\frac{D_0^S(3) (2 p_{12} m_S^2+p_{13} p_{23}) (p_{12} m_S^2+2 p_{13} p_{23})}{12p_{12}^3 p_{13}}
    -\frac{1}{12 p_{12} p_{13}} \ ,
\end{eqnarray}

\begin{eqnarray}
    f_{S7} &=& -\frac{B_0^S(1)}{2 p_{23}^3}
    -\frac{B_0^S(2) [p_{12} (2 p_{13}+p_{23})+p_{23} (3 p_{13}+p_{23})]}{4 p_{23}^3 (p_{12}+p_{23})^2}
    +\frac{B_0^S(4) (p_{12}+p_{13}+p_{23}) (2 p_{12}+3 p_{23})}{4 p_{23}^3 (p_{12}+p_{23})^2} \nonumber\\
    && -\frac{C_0^S(1) (2 p_{13}+p_{23}) (p_{23} m_S^2+p_{12} p_{13})}{4 p_{13} p_{23}^4} \nonumber\\
    && -\frac{C_0^S(2) \{m_S^2[p_{13} p_{23}^2+p_{12} p_{23} (2 p_{13}+p_{23})]
    +p_{12}^2 p_{13}(2 p_{13}+p_{23})\}}{4 p_{12}^2 p_{23}^4}
    -\frac{C_0^S(3) m_S^2}{4 p_{12}^2 p_{23}} \nonumber\\
    && +\frac{C_0^S(4) (p_{13}+p_{23}) [2 p_{23} m_S^2+p_{12} (2 p_{13}+p_{23})]}{4 p_{12} p_{23}^4} \nonumber\\
    && +\frac{C_0^S(5) [(2 p_{13}+p_{23}) p_{12}^4+2 p_{23} (m_S^2+2 p_{13}+p_{23}) p_{12}^3+p_{23}^2
    (3 m_S^2+2 p_{13}+p_{23}) p_{12}^2+m_S^2 p_{23}^4]}{4 p_{12}^2 p_{23}^4 (p_{12}+p_{23})} \nonumber\\
    && +\frac{C_0^S(6) m_S^2 (p_{12}+p_{13})^2}{4 p_{12}^2 p_{13} p_{23}^2}
    +\frac{D_0^S(1) [2 p_{23}^2 m_S^4+p_{12} p_{23} (8 p_{13}+3 p_{23}) m_S^2+2 p_{12}^2 p_{13}
    (2 p_{13}+p_{23})]}{4 p_{12} p_{23}^4} \nonumber\\
    && +\frac{D_0^S(2) (2 p_{13} m_S^4+p_{12} p_{23} m_S^2)}{4 p_{12} p_{13} p_{23}^2}
    +\frac{D_0^S(3) m_S^2 [2 p_{12} m_S^2+(p_{12}+2 p_{13}) p_{23}]}{4 p_{12}^2 p_{23}^2}
    +\frac{1}{4 p_{23}^2 (p_{12}+p_{23})} \ ,
\end{eqnarray}

\begin{eqnarray}
    f_{S13}&=& \frac{B_0^S(1) [p_{12}(6 p_{23}^2 p_{13}-8 p_{13}^3)-3 (4 p_{12}+p_{13}) p_{23} p_{13}^2+(4
    p_{12}+3 p_{13}) p_{23}^3]}{24 p_{13}^2 p_{23}^2 (p_{13}+p_{23})^2} \nonumber\\
    && +\frac{B_0^S(2) (p_{23}-8 p_{12})}{24 p_{12} p_{23}^2}
    +\frac{B_0^S(3) (4 p_{12}+p_{13})}{24 p_{12} p_{13}^2} \nonumber\\
    && +\frac{B_0^S(4) (p_{12}+p_{13}+p_{23}) [2 p_{12} (4 p_{13}^3+6 p_{23} p_{13}^2-3 p_{23}^2 p_{13}-2
    p_{23}^3)-p_{13} p_{23} (p_{13}+p_{23})^2]}{24 p_{12} p_{13}^2 p_{23}^2 (p_{13}+p_{23})^2} \nonumber\\
    && +\frac{C_0^S(1) p_{12} [-8 p_{12} p_{13}^3-3 (2 m_S^2+p_{13}) p_{23} p_{13}^2+(4 p_{12}+3
    p_{13}) p_{23}^3]}{24 p_{13}^3 p_{23}^3} \nonumber\\
    && +\frac{C_0^S(2) [-8 p_{13} p_{12}^3-3 (2 m_S^2+p_{13}) p_{23} p_{12}^2+p_{13} p_{23}^3]}{24
    p_{12}^2 p_{23}^3}
    +\frac{C_0^S(3) (4 p_{12}^3+3 p_{13} p_{12}^2+p_{13}^3) p_{23}}{24 p_{12}^2 p_{13}^3} \nonumber \\
    && +\frac{C_0^S(4)}{24 p_{13}^3p_{23}^3 (p_{13}+p_{23})} \{4 p_{12} (2 p_{13}^3-p_{23}^3) (p_{13}+p_{23})^2+3 p_{13} p_{23} [(p_{13}-p_{23})(p_{13}+p_{23})^3 \nonumber\\
    && +2 m_S^2 p_{13} (p_{13}^2+2 p_{23} p_{13}-p_{23}^2)]\}
    +\frac{C_0^S(5) (p_{12}+p_{23}) [p_{13}(8p_{12}^3-p_{23}^3)+3 (2 m_S^2+p_{13}) p_{23} p_{12}^2]}{24 p_{12}^2 p_{13} p_{23}^3} \nonumber\\
    && -\frac{C_0^S(6) (p_{12}+p_{13})^2 (4 p_{12}^2-p_{13} p_{12}+p_{13}^2)}{24 p_{12}^2 p_{13}^3} \nonumber\\
    && +\frac{D_0^S(1) [8 p_{12}^2 p_{13}^2+p_{12} (14 m_S^2+3 p_{13}) p_{23} p_{13}+m_S^2
    (2 m_S^2+3 p_{13}) p_{23}^2]}{12 p_{13} p_{23}^3} \nonumber\\
    && -\frac{D_0^S(2) [-2 p_{13}^2 m_S^4+p_{13} (4 p_{12}+3 p_{13}) p_{23} m_S^2+p_{12} (4 p_{12}+3
    p_{13}) p_{23}^2]}{12 p_{13}^3 p_{23}} \nonumber\\
    && +\frac{D_0^S(3) (m_S^2 p_{12}-p_{13} p_{23}) (2 p_{12} m_S^2+p_{13} p_{23})}{12
    p_{12}^2 p_{13} p_{23}}
    -\frac{p_{23}-2 p_{13}}{12 p_{13} p_{23} (p_{13}+p_{23})} \ ,
\end{eqnarray}
where $S$=$h_1^+$, $h_2^+$, $h_3^+$, $h_4^+$, $d_1^{++}$, $d_2^{++}$, $d_3^{++}$, $G_W^+$, $C_W^+$, $\bar{C}_W^+$, $G_Y^+$, $C_Y^+$, $\bar{C}_Y^+$, $G_Y^{++}$, $C_Y^{++}$, $\bar{C}_Y^{++}$.
In the above expressions, we have introduced the following defintions for the Passarino-Veltman scalar functions:
\[
B_0^X(1)\equiv B_0(2p_{12},m_X^2,m_X^2), \
B_0^X(2)\equiv B_0(2p_{13},m_X^2,m_X^2), \
B_0^X(3)\equiv B_0(2p_{23},m_X^2,m_X^2),
\]
\[
B_0^X(4)\equiv B_0(m_{V^0}^2,m_X^2,m_X^2), \
C_0^X(1)\equiv C_0(0,0,2p_{12},m_X^2,m_X^2,m_X^2), \
C_0^X(2)\equiv C_0(0,0,2p_{13},m_X^2,m_X^2,m_X^2),
\]
\[
C_0^X(3)\equiv C_0(0,0,2p_{23},m_X^2,m_X^2,m_X^2), \
C_0^X(4)\equiv C_0(0,2p_{12},m_{V^0}^2,m_X^2,m_X^2,m_X^2), \
C_0^X(5)\equiv C_0(0,2p_{13},m_{V^0}^2,m_X^2,m_X^2,m_X^2),
\]
\[
C_0^X(6)\equiv C_0(0,2p_{23},m_V^2,m_X^2,m_X^2,m_X^2), \
D_0^X(1)\equiv D_0(0,0,0,m_{V^0}^2,2p_{12},2p_{13},m_X^2,m_X^2,m_X^2,m_X^2),
\]
\[
D_0^X(2)\equiv D_0(0,0,0,m_{V^0}^2,2p_{12},2p_{23},m_X^2,m_X^2,m_X^2,m_X^2), \
D_0^X(3)\equiv D_0(0,0,0,m_{V^0}^2,2p_{13},2p_{23},m_X^2,m_X^2,m_X^2,m_X^2),
\]
where $X$ denotes the virtual particle circulating in the loop, and $p_{ij}\equiv p_i\cdot p_j$ with $i,j$=1, 2, 3.

\end{document}